%% file: main.tex
\documentclass{article}

\usepackage[english]{babel}

\usepackage[letterpaper,top=2cm,bottom=2cm,left=3cm,right=3cm,marginparwidth=1.75cm]{geometry}
\usepackage[utf8]{inputenc}
\usepackage{textcomp}
\usepackage{subcaption} 
\usepackage{multirow}
\usepackage[normalem]{ulem}
\usepackage{amsmath}
\usepackage{graphicx}
\usepackage{enumitem}
\usepackage[left]{lineno}
\usepackage[colorlinks=true, linkcolor=blue, citecolor=blue, urlcolor=blue]{hyperref}
\usepackage{authblk} 

\title{A Low-energy Threshold and Multi-messenger Trigger System for the JUNO Experiment}
\input{"Authors_v2"}

\begin{document}
% \linenumbers
\maketitle

\begin{abstract}
The Jiangmen Underground Neutrino Observatory (JUNO) is a 20-kiloton liquid scintillator neutrino detector, located 650 meters (1800 m.w.e.) underground in Jiangmen, Guangdong, China. JUNO is primarily designed for reactor neutrino measurements and has been taking data since 2025. With the largest mass of its kind and an excellent energy resolution, JUNO is a leading observatory for high-precision measurements of MeV neutrinos. 
The standard global trigger system serves as the primary trigger for JUNO. We present a newly developed multi-messenger trigger system that extends the capabilities of the global trigger by providing a lower energy threshold and an independent monitoring capability. During the 2025 operation, it achieved an effective energy threshold of approximately $110 \pm 10$~keV, providing a lower threshold configuration suitable for low-energy event analysis.
The system shows the potential to further reduce the threshold to well below 100~keV. Based on the multi-messenger trigger system, an astrophysical monitor has been developed to receive and process external alerts from other messengers, such as gravitational-wave observations. A Transient Neutrino Burst Monitor is integrated to detect short-time-scale neutrino burst events and enables real-time monitoring of transient astrophysical phenomena. The system is sensitive to neutrino bursts from core-collapse supernovae within a distance of about 250~kpc.

\end{abstract}
\tableofcontents
\section{Multi-messenger Astronomy}
\label{Introduction}

Within the past decade, the detection of gravitational waves~\cite{Abbott2016} and the discovery of an extra-galactic high-energy neutrino flux~\cite{Aartsen2013, Aartsen2014} have opened a new era of multi-messenger astronomy. Extreme cosmic events such as core-collapse supernovae and neutron star mergers are expected to produce multiple messengers—gravitational waves, neutrinos, electromagnetic radiation, and cosmic rays—within a short time span. 

SN1987A has been a major milestone in neutrino astronomy, when a handful of MeV-scale neutrinos were detected by Kamiokande-II \cite{Hirata1987}, IMB~\cite{Haines1988}, and Baksan~\cite{Alexeyev1988} several hours before the optical signal~\cite{Kunkel1987}. This discovery demonstrated that neutrinos escape promptly from the collapsing core, initiated a new observational era, and even constrained fundamental properties of the electron antineutrino~\cite{Arnett1989}. Future galactic supernovae are expected to deliver unprecedented statistics: $\sim99\%$ of the gravitational binding energy is carried away in neutrinos over $\sim$10 s~\cite{Woosley2005, Janka2007, Janka2012}. Models of the neutrino emission predict a short ($30$–$50$ ms) neutronization burst~\cite{Kachelriess2005}, followed by the intense accretion and cooling phases of the proto-neutron star, and possible additional bursts from quark–hadron phase transitions~\cite{Sagert2009, Gentile1993, Fischer2011}. In addition, a detectable pre-supernova neutrino flux is expected from late stellar burning stages~\cite{Odrzywolek2004}.

Core-collapse supernovae occur in the Milky Way at an estimated rate of 1–3 per century~\cite{Cappellaro1993, Adams2013, Rozwadowska2020},
making them the most promising nearby sources of astrophysical neutrinos. The next galactic core-collapse supernova is expected to be observed with unprecedented statistics by current and next-generation neutrino detectors such as Super-Kamiokande~\cite{Fukuda2003}, IceCube~\cite{Achterberg2006}, KM3NeT~\cite{KM3NeT2021CCSN}, DUNE~\cite{Acciarri2016DUNE}, Hyper-Kamiokande~\cite{Abe2018HyperK} and JUNO~\cite{JUNO2021Review}. 

Among them, JUNO, which began physics data taking in August 2025~\cite{JUNO_Initial_Performance_2025, JUNO_ReactorOscillation_2025}, is a 20-kton liquid-scintillator detector and the largest of its kind to date. Its excellent energy resolution and wide sensitive energy range make JUNO one of the most powerful experiments for supernova neutrino observation. JUNO is capable of detecting all flavors of post-shock neutrinos in the $\mathcal{O}(10)\mathrm{MeV}$ energy range thanks to its sensitivity to multiple interaction channels, including inverse beta decay, neutrino--proton and neutrino--electron elastic scattering. 
For a typical galactic distance of 10 kpc and representative supernova parameters, JUNO is expected to record approximately 7000 events for a 20 $M_{\odot}$ progenitor~\cite{Migenda2021} (including all interaction channels based on the Nakazato model~\cite{nakazato2013} and a characteristic neutrino luminosity of order $\sim 10^{53}$~erg/s).
The high statistics across multiple interaction channels uniquely position JUNO to provide precise measurements of the time, spectral, and flavor evolution of supernova neutrinos.

For binary neutron star mergers, gravitational waves are predicted to precede neutrinos by tens of milliseconds during the in-spiral phase~\cite{Perego2014}, whereas in core-collapse supernovae the two messengers are released nearly simultaneously under certain conditions~\cite{Mueller2004}. Both phenomena are expected to emit neutrinos with luminosities of order $\sim 10^{53}$ erg/s, with typical neutrino energies of $\sim 10$--$20$~MeV as predicted by simulations~\cite{Cusinato2022BNS}, though the spectral details remain uncertain~\cite{Sekiguchi2011}. Galactic merger rates, however, are very low ($10^{-6}$–$10^{-4}$ yr$^{-1}$)~\cite{Burgay2003, Abbott2017_GW170817}, and searches for neutrino counterparts of GW170817 yielded null results, likely due to geometric collimation or distance limitations~\cite{Albert2017}.

This paper reports on a new multi-messenger trigger system designed for JUNO to maximize its astrophysics potential. During the 2025 operation, the system achieved an effective energy threshold of approximately $110 \pm 10$~keV, with the capability to be further reduced to well below 100~keV. This low threshold enables JUNO to perform broadband monitoring of the transient neutrino sky. Furthermore, it provides a powerful tool for exploring low-energy physics.

The content of this paper is arranged as follows. Section~\ref{sec:juno_detector} briefly introduces the JUNO detector and trigger systems. The multi-messenger trigger system is presented in Section~\ref{sec:Multi-messenger_Trigger_System}, organized in two subsections: Section~\ref{sec:Low_Energy_Threshold_Trigger} presents the trigger algorithms, hardware architecture, data-taking status in 2025, and trigger efficiency estimation. And Section~\ref{sec:astrophysical_monitor} introduces the astrophysical monitor, including the Transient Neutrino Burst Monitor based on the trigger and the External Alert Processor, which receives and processes alerts from external observatories and other multi-messenger facilities. The performance and potential of the JUNO multi-messenger trigger system for core-collapse supernova neutrinos are presented in Section~\ref{sec:ccsn}. Finally, a summary and discussion is given in Section~\ref{sec:Conclusion}.

\section{The JUNO Detector and Trigger Systems}
\label{sec:juno_detector}

JUNO is the largest liquid scintillator (LS) detector of the coming decade, with a 20-kton active target located 650 m underground in Jiangmen, Guangdong, China. One of its primary scientific goals is to determine the neutrino mass ordering through precision measurement of reactor neutrino oscillations at a medium baseline of about 53 km from the Yangjiang and Taishan Nuclear Power Plants~\cite{An2016}.

\begin{figure}[htbp]
  \centering
  \includegraphics[width=0.7\textwidth]{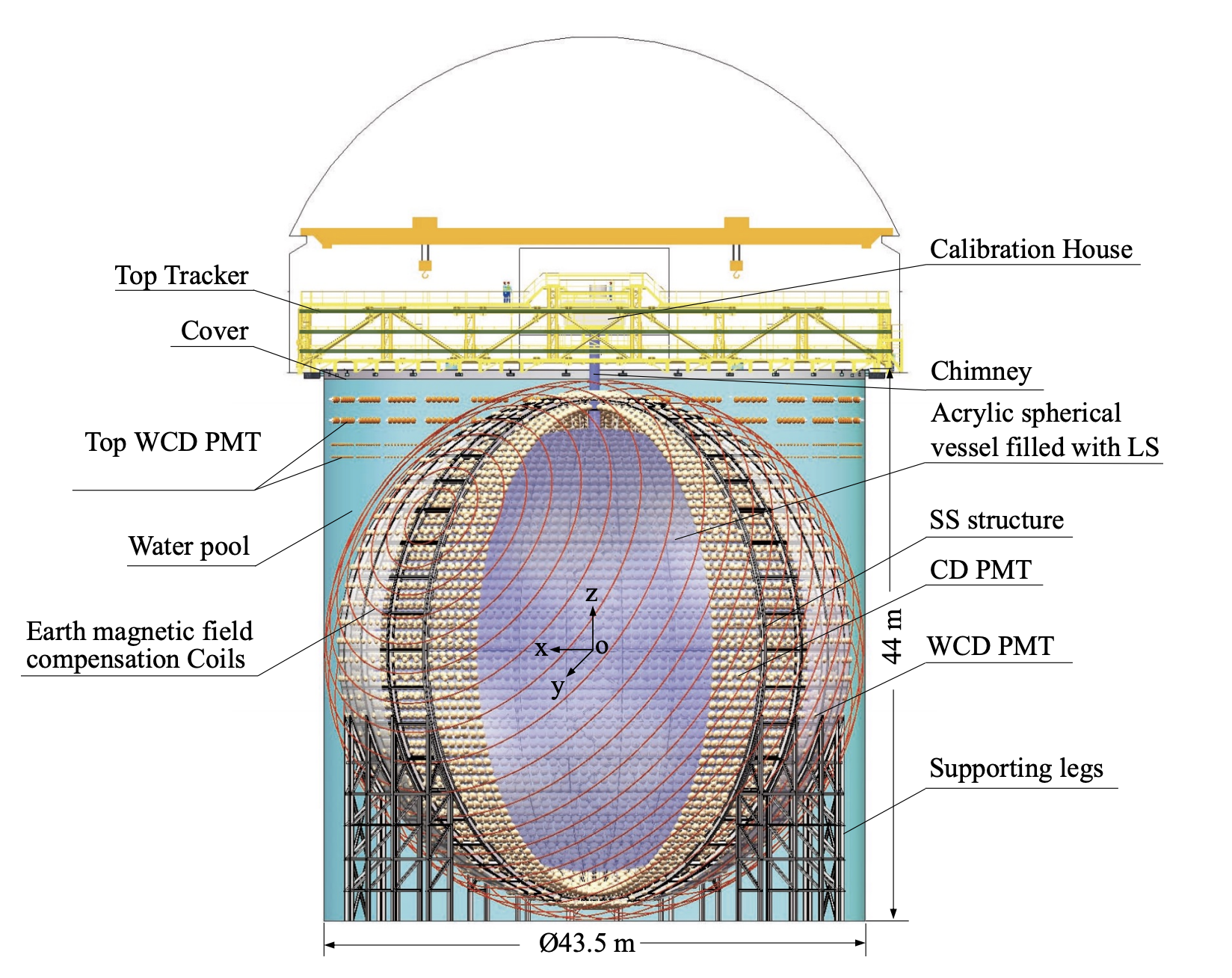}
  \caption{Schematic view of the JUNO detector~\cite{JUNO_Initial_Performance_2025}}
  \label{fig:JUNO_structure}
\end{figure}

The Central Detector (CD) consists of a 35.4~m diameter acrylic sphere filled with 20~kton LS, instrumented with 17,596 high-quantum-efficiency 20-inch PMTs and 25,587 3-inch PMTs, providing over 75\% photocathode coverage. This configuration yields a photoelectron collection of about 1785~PE/MeV at the detector centre~\cite{JUNO_Initial_Performance_2025}, corresponding to a measured energy resolution of about 3.4\% at 1~MeV scale, a critical requirement for resolving the neutrino mass ordering. Surrounding the CD is a cylindrical Water Pool (WP) equipped with 3352 PMTs to function as a water Cherenkov detector. It serves simultaneously as a cosmic muon veto and as shielding against external radioactivity. On top of the water pool, a plastic scintillator Top Tracker (TT) further improves muon tracking capabilities. A schematic view of the detector layout is shown in Figure~\ref{fig:JUNO_structure}. Further technical details can be found in Refs.~\cite{JUNO2021Review, JUNO_Initial_Performance_2025}. In standard data-taking operation, JUNO employs a global trigger system as the primary event selection mechanism for data acquisition. The global trigger is designed to efficiently collect physics events while suppressing detector noise and uncorrelated backgrounds. In the current configuration, the global trigger corresponds to an effective energy threshold of approximately 200~keV, below which the trigger efficiency rapidly decreases. This threshold is optimized for reactor neutrino measurements and general physics analyses.

\begin{figure}[htbp]
  \centering
  \includegraphics[width=0.55\textwidth]{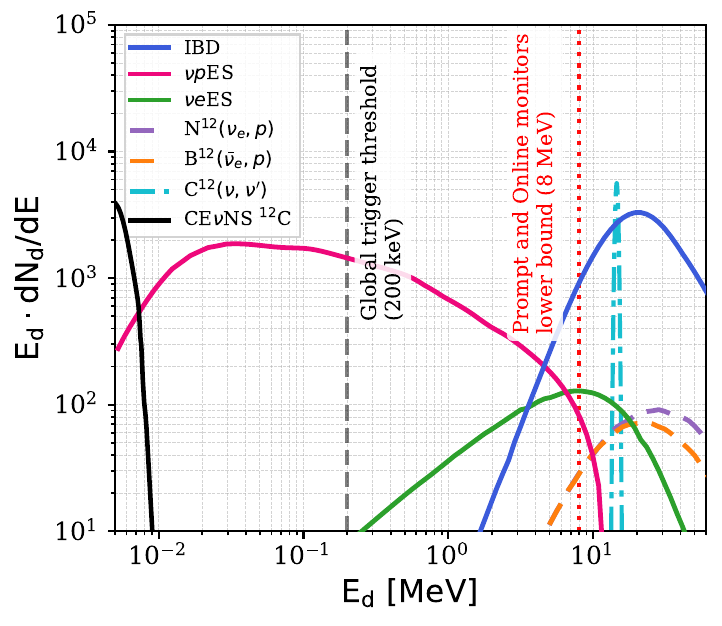}
\caption{Visible energy ($\mathrm{E_d}$) spectra of different interaction channels in JUNO for a $30\,M_\odot$ core-collapse supernova (Nakazato model~\cite{nakazato2013}). The dashed vertical line indicates the effective energy threshold of the global trigger at approximately 200~keV. The dotted vertical line shows the lower energy bound of the Prompt and Online monitors at about 8~MeV.}
  \label{fig:energy_spectrum}
\end{figure}

Two systems, the Prompt Monitor and the Online Monitor, have been developed within JUNO to perform real-time searches for core-collapse supernova (CCSN) neutrino bursts~\cite{Abusleme2024SNMonitoring}. The Prompt Monitor is implemented on the global-trigger board, while the Online Monitor is based on the data acquisition (DAQ) system and the Online Event Classification (OEC) framework. Early detection of CCSN neutrinos enables a fast warning of the explosion and is essential for low-latency multi-messenger observations. 
Figure~\ref{fig:energy_spectrum} shows the visible energy spectra of the different interaction channels in JUNO, which are summarized in Table~\ref{tab:interaction_channels}, expected for the $30\,M_\odot$ Nakazato model~\cite{nakazato2013}.

To ensure a clean background, the Prompt Monitor and the Online Monitor operate with a low-energy threshold of about 8~MeV and use the IBD channel as the primary signal. However, this choice inevitably suppresses a large fraction of the $\nu$pES and $\nu$eES signals. These channels still provide significant statistics above the global trigger threshold, while even lower-energy interactions such as CE$\nu$NS lie entirely below the current detection range.

\begin{table}[htbp]
    \centering
    \renewcommand{\arraystretch}{1.3} % 稍微增加行距，防止公式粘连
    \begin{tabular}{l p{0.55\textwidth} c}
        \hline
        \hline
        \textbf{Channel} & \textbf{Process} & \textbf{Visible Energy Range} \\
        \hline
        IBD & Inverse Beta Decay ($\bar{\nu}_e + p \rightarrow e^+ + n$) & $\mathcal{O}(10)$~MeV \\
        $\nu$pES & Neutrino-Proton Elastic Scattering ($\nu + p \rightarrow \nu + p$) & sub-MeV \\
        $\nu$eES & Neutrino-Electron Elastic Scattering ($\nu + e^- \rightarrow \nu + e^-$) & sub-MeV to $\mathcal{O}(10)$~MeV \\
        ${}^{12}\mathrm{C}(\nu_e, e^-){}^{12}\mathrm{N}$ & $\nu_e$ Charged-Current on carbon ($\nu_e + {}^{12}\mathrm{C} \rightarrow e^- + {}^{12}\mathrm{N}$) & $\mathcal{O}(10)$~MeV \\
        ${}^{12}\mathrm{C}(\bar{\nu}_e, e^+){}^{12}\mathrm{B}$ & $\bar{\nu}_e$ Charged-Current on carbon ($\bar{\nu}_e + {}^{12}\mathrm{C} \rightarrow e^+ + {}^{12}\mathrm{B}$) & $\mathcal{O}(10)$~MeV \\
        ${}^{12}\mathrm{C}(\nu, \nu')$ & Neutral-Current excitation ($\nu + {}^{12}\mathrm{C} \rightarrow \nu + {}^{12}\mathrm{C}^*$) & 15.11~MeV ($\gamma$) \\
        CE$\nu$NS & Coherent Elastic Neutrino-Nucleus Scattering ($\nu + {}^{12}\mathrm{C} \rightarrow \nu + {}^{12}\mathrm{C}$) & sub-MeV \\
        \hline
        \hline
    \end{tabular}
    \caption{Summary of supernova neutrino interaction channels in JUNO.}
    \label{tab:interaction_channels}
\end{table}

This observation motivates the development of a lower-threshold trigger system, the multi-messenger (MM) trigger. In addition to providing prompt CCSN alerts, the MM trigger is designed to receive and process external alerts from other experiments and messengers, which motivates its name.
Beyond rare CCSN events, the lowered energy threshold potentially enables the study of sub-MeV neutrino physics, including pp-chain solar neutrinos~\cite{Borexino_ppchain_2018}, which probe the dominant energy production mechanism in the Sun. The enhanced low-energy sensitivity of the MM trigger is also particularly important for searches for neutrino magnetic moments ($\nu$MM)~\cite{Ye2021_NeutrinoMagMoment_LSdetectors}, which can be probed via $\nu$eES. A non-zero $\nu$MM introduces an additional contribution to the cross section that scales as $1/T$, where $T$ is the electron recoil energy, leading to a strong enhancement at low energies. By significantly lowering the detection threshold, the MM trigger provides a unique opportunity to probe such effects in both steady solar neutrino fluxes and transient CCSN signals.

The MM trigger system has been integrated into the JUNO data acquisition flow since the commissioning stage. 
In this work, we use real trigger rates together with simulations to estimate the achievable energy threshold of the MM trigger system, and we present the performance of the newly developed monitor based on early data collected in 2025.
These first results demonstrate the capability of the MM trigger system, and as JUNO continues stable operation, more comprehensive and increasingly precise results are expected in the future.

\section{Multi-messenger Trigger System}
\label{sec:Multi-messenger_Trigger_System}

\subsection{Low Energy Threshold Trigger}
\label{sec:Low_Energy_Threshold_Trigger}

Lowering the energy threshold to the sub-MeV regime introduces significant challenges from detector backgrounds, which are dominated by PMT dark noise at the raw data level. To suppress these backgrounds, we develop a likelihood-based trigger algorithm that exploits the spatial non-uniformity of physical signals in contrast to the uniform distribution of dark noise. The algorithm is implemented on a FPGA and integrated into the JUNO data acquisition chain for real-time processing.

The performance of the MM trigger is evaluated using early data collected in 2025 by comparing its trigger rate with the global trigger. Based on trigger rate and simulations, we estimate that the MM trigger can achieve an effective energy threshold of approximately $110 \pm 10$~keV. The feasibility of the low-energy performance is further studied using simulations of dedicated calibration sources.

\subsubsection{Low Energy Background}
\label{sec:Background}
In the energy region around and below 200 keV, the dominant noise sources include the dark noise of the 20-inch PMTs, the $\beta$-decay of $^{14}$C in the LS, and radioactive isotopes in the PMT glass, in order of importance by their contributions to the data bandwidth.
\begin{itemize}
    \item \textbf{Dark noise}: Thermal emission and spontaneous electron emission inside the PMTs cause random single photoelectron signals even in the absence of light. There are 17,596 20-inch PMTs in the CD, and the average dark noise rate is roughly 20 kHz, which would generate an enormous amount of data. Fortunately, the dark noise is spatially uniform, allowing the trigger algorithm to filter out the vast majority of such random signals effectively.
    \item \textbf{$\beta$-decay of $^{14}$C in LS}: Naturally occurring $^{14}$C in the liquid scintillator undergoes beta decay, producing low-energy electrons with a maximum energy of 156 keV. It is the dominant source of physical background around $\sim 100$~keV.
    \item \textbf{Radioactive isotopes in PMT glass}: $^{238}\mathrm{U}$ and $^{232}\mathrm{Th}$ present in the LPMT glass~\cite{JUNO2021Review} undergo radioactive decay, emitting beta or gamma particles that can produce Cherenkov light in the surrounding water. Due to the intrinsically low light yield of Cherenkov radiation and the fact that these events originate at the detector edge, their effective visible energy is typically below $\sim 100$~keV.
\end{itemize}

At the raw data level, PMT dark noise is by far the dominant background source. Even without storing full waveforms and retaining only the basic LPMT charge and timing information, the data bandwidth is $\sim$8~GB/s. Such a high data rate makes it impractical to record all signals and necessitates an efficient real-time trigger system. To suppress this background, we develop a likelihood-based trigger algorithm to suppress the dominant dark noise contribution by exploiting its spatial uniformity. Before introducing the trigger algorithm in detail, we first describe the electronic hardware system that enables real-time data collection and processing in JUNO.

\subsubsection{Electronic Hardware System}
\label{sec:Electronic_Hardware_System}

\begin{figure}[htbp]
  \centering
  \includegraphics[width=0.55\textwidth]{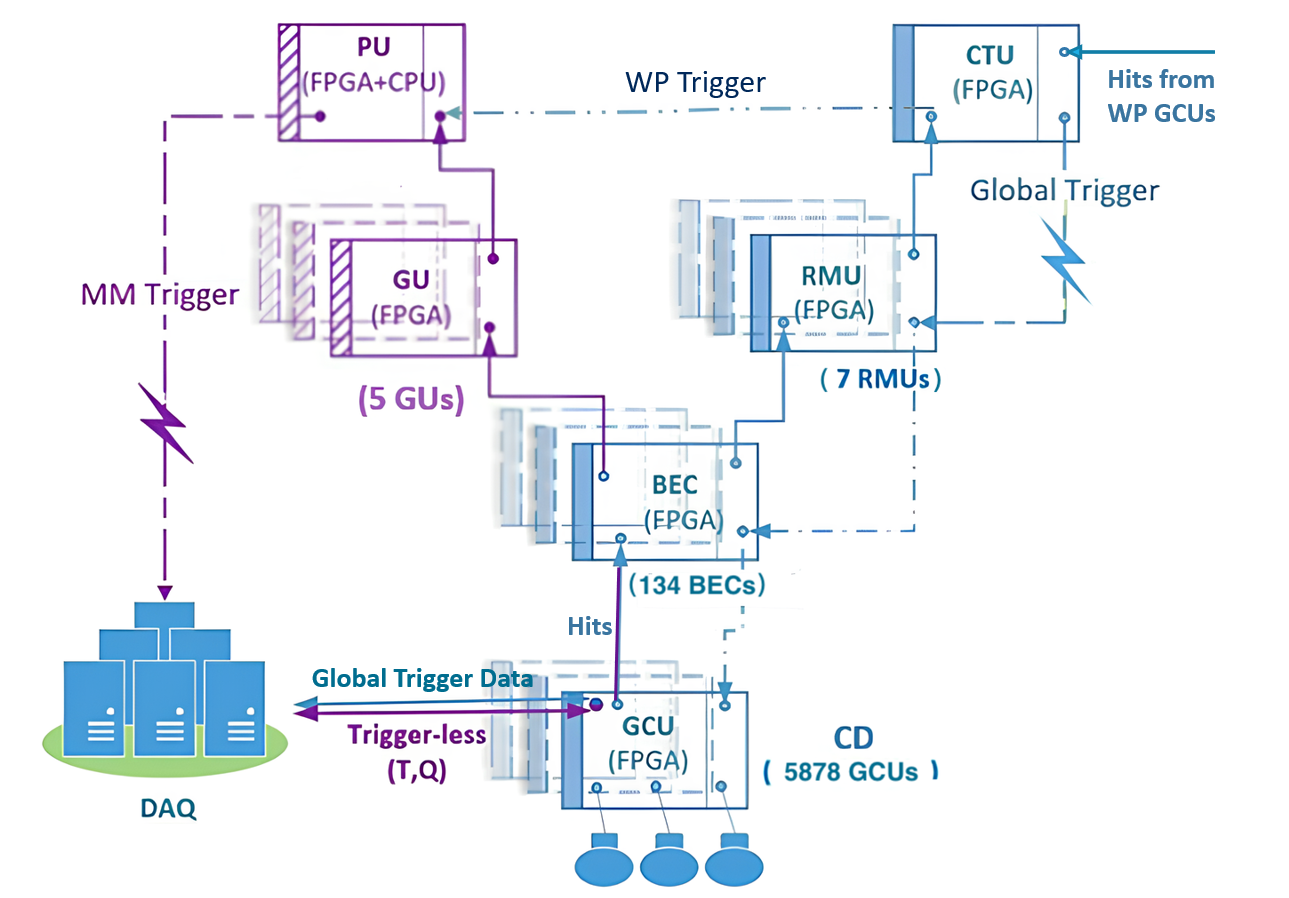}
  \caption{Schematic diagram of the JUNO trigger electronics system. The standard global trigger data flow is shown in blue, while the MM trigger data flow is indicated in purple.}
  \label{fig:hardware_structure}
\end{figure}

The JUNO CD is instrumented with 17,596 20-inch PMTs and is read out by a hierarchical electronic system, as illustrated in Figure~\ref{fig:hardware_structure}. In the standard JUNO electronics chain, the analog PMT signals are first digitized by 5,878 Global Control Units (GCUs)~\cite{Cerrone2023NIMA, Coppi2023Mass}, each connected to three PMTs. The GCUs extract basic hit information by applying a threshold to the digitized signals. A hit is defined as a summed signal above the threshold from up to three PMTs connected to a GCU, computed every 16~ns (one clock cycle). These hits are then transmitted to the Back-End Cards (BECs), which forward the digitized hit information to the Reorganize and Multiplex Units (RMUs)~\cite{Aloisio2017RMU}. The RMUs aggregate the data and send them to the Central Trigger Unit (CTU), where the global trigger is formed. In addition to the CD PMTs, the CTU also receives hit information from the WP PMTs, providing additional inputs for the global trigger formation. The global trigger decision is then distributed back to all GCUs, initiating a synchronized readout of PMT data, which are subsequently transmitted to the DAQ system for event building.

Building upon this standard infrastructure, the MM trigger system introduces an additional processing chain. Instead of relying solely on the CTU, the MM system receives hit information from the BECs and routes it to the Gather Units (GUs). Each GU integrates a White Rabbit (WR) timing system~\cite{lipinski2011white}, providing timestamps with 8~ns resolution for precise synchronization. The time-aligned hit data are then transmitted to the Processing Unit (PU), where the trigger algorithms are executed in real time. The PU generates trigger decisions and sends the corresponding trigger information to the DAQ system. In parallel, the DAQ continuously receives trigger-less PMT information (time and charge, TQ) from the GCUs. Based on the MM trigger decisions, the DAQ selectively builds events from the TQ data stream. The firmware architectures of PU and GU are presented in Appendix~\ref{appendix:Firmware}.

\begin{figure}[htbp]
    \centering
    \begin{subfigure}{0.40\textwidth}
        \centering
        \includegraphics[width=\textwidth]{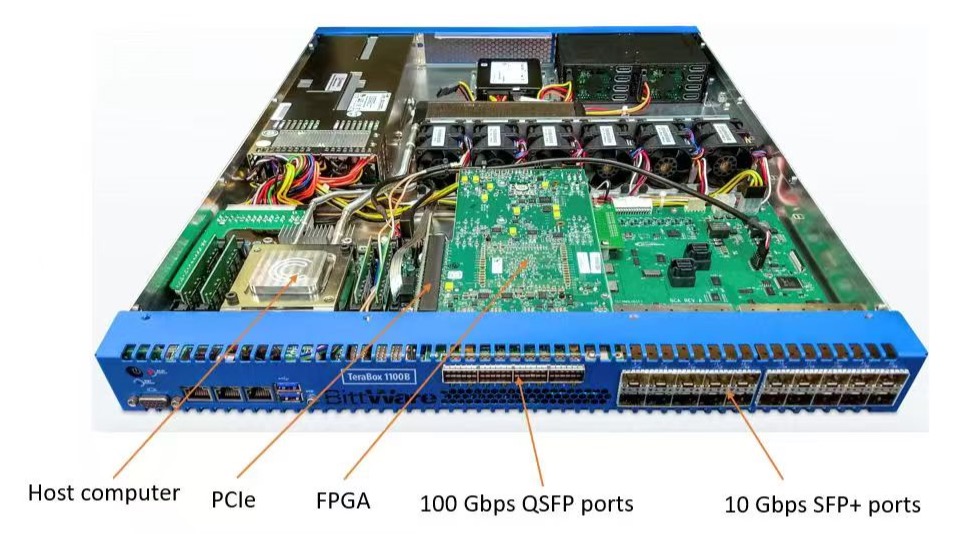}
        \caption{Photograph of the PU TeraBox system.}
    \end{subfigure}%
    \hfill
    \begin{subfigure}{0.53\textwidth}
        \centering
        \includegraphics[width=\textwidth]{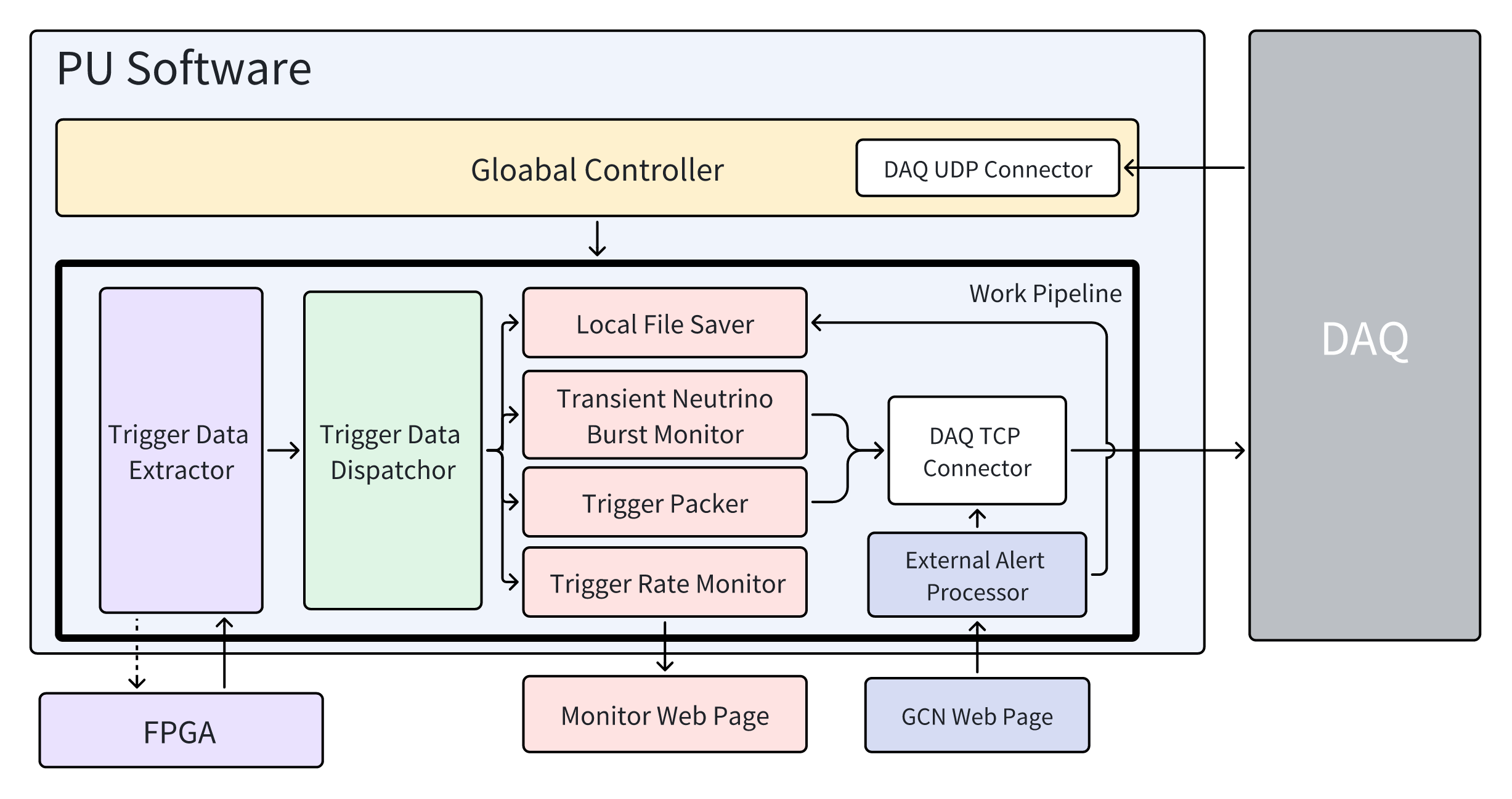}
        \caption{Illustration of the PU software.}
    \end{subfigure}
    \caption{Overview of the PU system.}
    \label{fig:PU}
\end{figure}

The PU, which serves as the core of the MM trigger system, is implemented on a TeraBox-1100L FPGA PCIe server~\cite{terabox1100l}, as shown in Figure~\ref{fig:PU}(a), equipped with a Xilinx UltraScale+ FPGA and a host CPU. The likelihood-based trigger algorithms described in Section~\ref{sec:Trigger_Algorithms} are deployed on the FPGA to enable fast, real-time processing. In parallel, a data link from the CTU provides WP global trigger information to the PU, which is used for muon veto in the monitor.

Figure~\ref{fig:PU}(b) illustrates the host software architecture of the PU. 
It continuously extracts trigger information from the FPGA and forwards it to the DAQ system, while also sending slow control commands, such as trigger threshold settings, back to the FPGA. Benefiting from a multi-threaded architecture, the trigger data is processed in parallel by different functional modules for real-time monitoring and data handling. They can generate and send alert messages to the DAQ. 
In addition to this primary functionality, the software further provides services including local file storage, external alert processing, and trigger rate monitoring.

\subsubsection{Data Acquisition System for MM trigger}
\label{sec:DAQ}

The DAQ system reads out the full detector data over the network. 
The data from individual detectors (e.g., CD and WP) are segmented and assembled at fixed time intervals to form time fragments, which serve as the basic units for online processing. 
Online processing is performed on a per-time-fragment basis~\cite{Chen2025JUNODAQ}. 

\begin{figure}[htbp]
  \centering
  \includegraphics[width=0.7\textwidth]{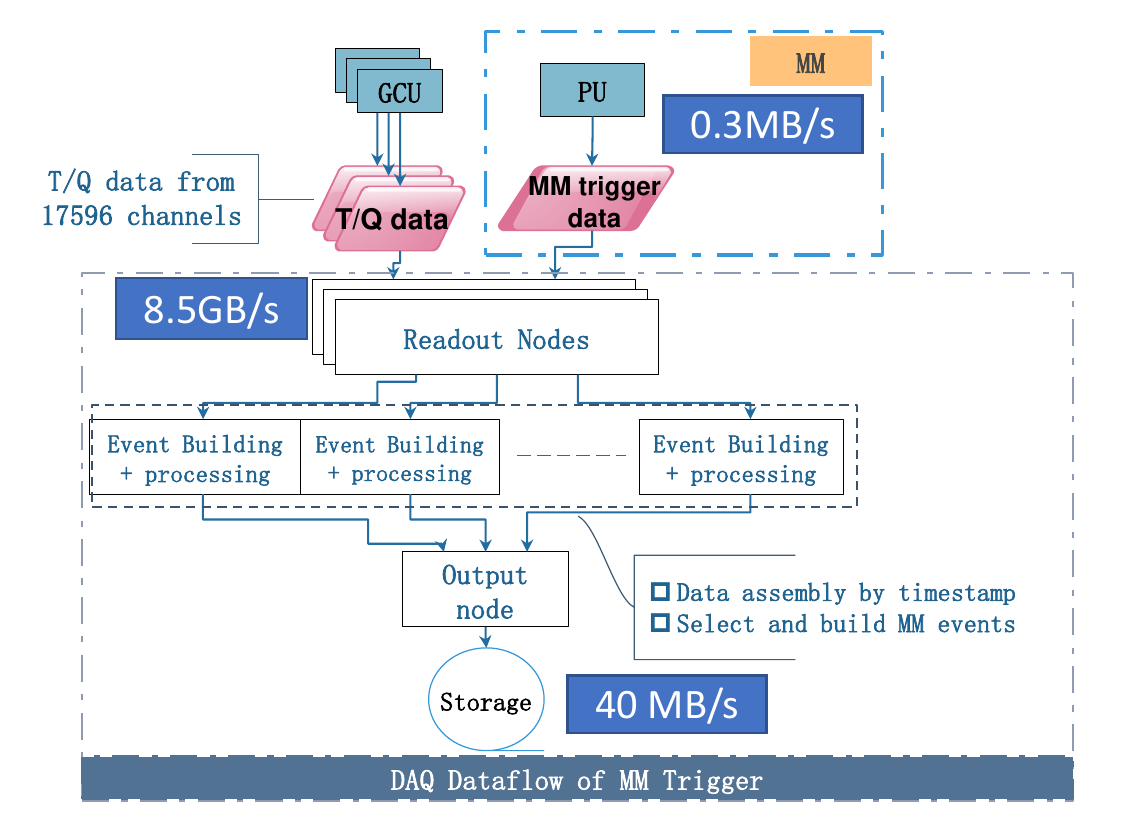}
  \caption{The schematic design of data acquisition system for the MM trigger of JUNO.
}
  \label{fig:DAQ}
\end{figure}

For MM-triggered events, as illustrated in Figure~\ref{fig:DAQ}, the raw T/Q data stream has a bandwidth of approximately 8.5~GB/s. 
Guided by the timing information provided by the MM trigger, the DAQ system selects T/Q data within a predefined time window (700~ns, from $t_{\mathrm{trigger}}-300$~ns to $t_{\mathrm{trigger}}+400$~ns) and assembles them into MM events. 
% After event building, the data are stored locally on site, while a fraction of about 1\% is transferred to the IHEP data centre for fast analysis.
After event building, the data is stored locally on site, while a fraction of about 1\% is transferred to the IHEP data centre for immediate access and analysis by the collaboration.

\subsubsection{Trigger Algorithm}
\label{sec:Trigger_Algorithms}

This section describes the trigger algorithms used in JUNO, i.e., the standard global trigger and the likelihood-based MM trigger.

\paragraph{Global trigger.}
The standard JUNO global trigger is formed in the CTU based on a multiplicity algorithm. The hits from all PMTs are summed within a fixed 304~ns time window, and a trigger is issued when the total multiplicity exceeds a threshold of about 350~hits~\cite{JUNO_Initial_Performance_2025}, which corresponds to an energy of approximately 200~keV.

\paragraph{Likelihood-based MM trigger.}
To achieve a lower energy threshold, we introduce a likelihood-based trigger algorithm. The key idea is to exploit the difference between the spatially uniform distribution of dark noise and the clustered topology of physical events. A trigger time window of 192~ns is adopted, chosen as a compromise between suppressing dark noise and preserving signal clustering in space and time. Within each time window, the 17,596 PMTs are divided into a space--time grid with $B_t \times B_\theta \times B_\phi = 4 \times 4 \times 4$ bins. Here, $B_t = 4$ means the 192~ns trigger time window is equally divided into four 48~ns temporal bins. Because the JUNO central detector is spherical, the position of each PMT is naturally defined by its polar angle $\theta$ and azimuthal angle $\phi$. Therefore, the spatial term $B_\theta \times B_\phi = 4 \times 4$ represents a uniform binning of PMT positions in $\cos\theta$ and $\phi$ across the entire sphere. The likelihood is constructed as the product of Poisson probabilities over all 64 bins:
\begin{equation}
    \mathcal{L} = \prod_{i=1}^{4} \prod_{j=1}^{4} \prod_{k=1}^{4} \frac{\lambda^{N_{ijk}}}{N_{ijk}!} e^{-\lambda},
    \label{likelihood_eq}
\end{equation}
where $N_{ijk}$ denotes the number of hits in bin $(i,j,k)$, and $\lambda$ is the expected number of hits per bin for pure dark noise, with a typical value of $\lambda = 1.58$.

\begin{figure}[htbp]
  \centering
  \includegraphics[width=0.5\textwidth]{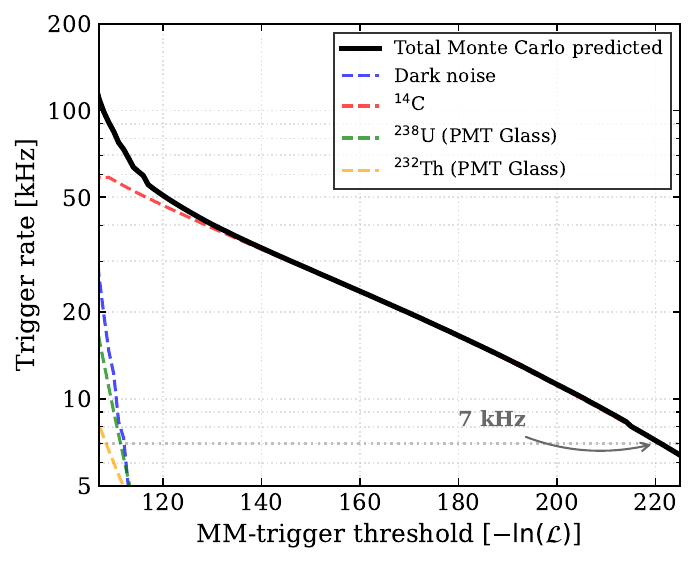}
  \caption{MM-trigger rate as a function of the likelihood threshold for different background components from simulation.}
  \label{fig:mm_threshold_scan}
\end{figure}

In practice, the FPGA in the PU evaluates a scaled negative log-likelihood, $-\ln \mathcal{L}$, using pre-calculated lookup tables. A trigger is issued when the summed likelihood value exceeds a predefined threshold.

\begin{figure}[htbp]
  \centering
  \includegraphics[width=0.7\textwidth]{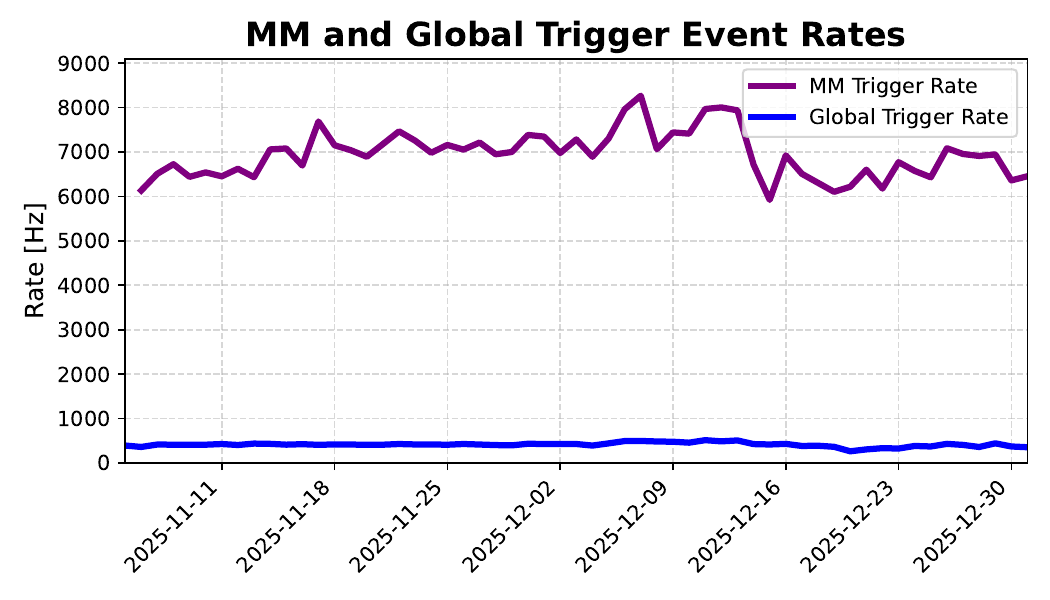}
  \caption{The comparison between MM trigger rate and Global trigger rate during data taking.
}
  \label{fig:mmctu_triggerrate}
\end{figure}

The dependence of the MM-trigger rate on the likelihood threshold is shown in Figure~\ref{fig:mm_threshold_scan}. Dark noise and radioactive backgrounds from PMT glass are efficiently suppressed with increasing threshold, while $^{14}$C decays dominate the trigger rate for $-\ln \mathcal{L}$ thresholds larger than 120. 
Figure~\ref{fig:mmctu_triggerrate} shows the trigger rates observed during the 2025 data-taking period with a MM-trigger $-\ln\mathcal{L}$ threshold of 218.
The global trigger rate is about 400~Hz, while the MM-trigger rate is around 7~kHz. Comparing simulation and data, we find that the observed MM-trigger rate is predominantly due to $^{14}$C decays.

\subsubsection{Trigger Threshold}
\label{sec:Trigger_Efficiency}

\begin{figure}[htbp]
  \centering
  \includegraphics[width=0.55\textwidth]{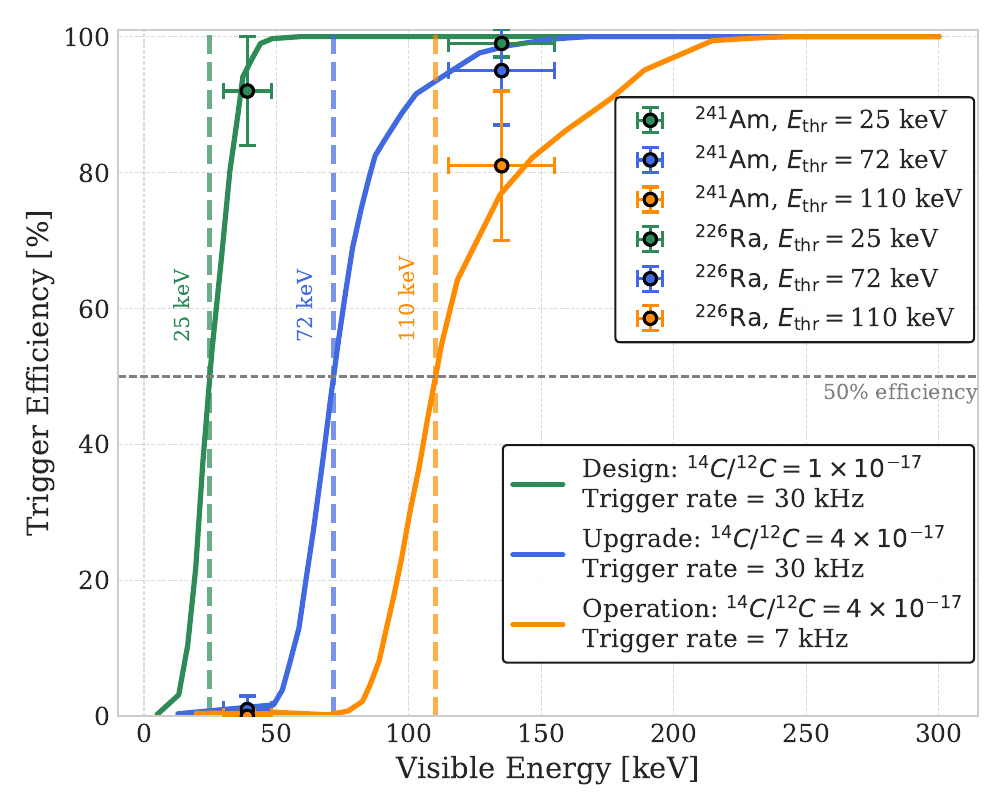}
  \caption{The trigger efficiency as a function of electron visible energy for different $^{14}$C abundances and trigger rate configurations. The markers represent the simulated response of the calibration sources ($^{241}$Am and $^{226}$Ra) at these specific thresholds. The horizontal error bars account for the uncertainty in the deposition process from gamma energy to visible energy, while the vertical error bars represent the uncertainty associated with the event reconstruction algorithm.}
  \label{fig:trigger_eff_new}
\end{figure}

Considering the observed MM-trigger rate of about 7~kHz, which is predominantly driven by $^{14}$C decays, we use a representative $^{14}$C/$^{12}$C ratio of $4.0 \times 10^{-17}$ to constrain the trigger operating condition. 
This value is chosen as the central value of the range $(3-5) \times 10^{-17}$ reported in the JUNO initial performance results~\cite{JUNO_Initial_Performance_2025}.
Under this condition, the trigger efficiency is evaluated using the official JUNO simulation software~\cite{lin2023junosw}, as shown in Figure~\ref{fig:trigger_eff_new}. In the simulation, electrons are generated uniformly throughout the full liquid scintillator volume, with energies sampled in the range of 0--300~keV and isotropic initial directions. 
Nominal PMT dark noise, with an average dark count rate of about 19 kHz, is included to reproduce realistic detector conditions.
The effective MM-trigger threshold is defined as the visible energy corresponding to 50\% trigger efficiency. Based on the simulation, this value is approximately $110 \pm 10$~keV, where the uncertainty mainly comes from the detector-position dependence, as indicated by the orange curve corresponding to the operating condition.

For the initial design of the MM trigger, assuming a total trigger rate of 30 kHz and a $^{14}$C/$^{12}$C ratio of $1.0 \times 10^{-17}$, the simulations indicate an effective energy threshold of about 25 keV (green curve). Under these conditions, the 30 kHz budget is composed of approximately 22 kHz from $^{14}$C decay and 8 kHz from the combined contribution of PMT dark noise and radioactive isotopes in the PMT glass. 

If the JUNO acquisition system can be upgraded in the coming years to operate at a total trigger rate of 30~kHz, for example through increased data bandwidth and DAQ throughput, the energy threshold could be reduced to approximately 72~keV, as illustrated by the blue curve in Figure~\ref{fig:trigger_eff_new}.
This indicates that even in the presence of a higher-than-expected $^{14}$C concentration, the MM trigger would still enable the recording of events well below 100~keV, maintaining a significant advantage in low-energy sensitivity compared to the Global Trigger. However, it should be noted that successfully triggering on these low-energy events does not guarantee their successful detection in offline analysis, as the substantial background must still be mitigated.

\subsubsection{Calibration Methodology}
\label{sec:calib}

Because the actual in-situ low-energy calibration campaigns are still being planned, this section presents a simulation-based feasibility study to demonstrate how the true trigger efficiency will be calibrated in the future. To achieve the excellent energy resolution required by JUNO, multiple radioactive sources have been deployed for detector calibration \cite{Abusleme2021Calibration}. Since the MM trigger operates at a low energy threshold, dedicated low-energy calibration sources are needed to characterize and validate the trigger efficiency (see Fig.~\ref{fig:trigger_eff_new}). Following the procedure of Ref.~\cite{Takenaka2024}, a $^{241}\mathrm{Am}$ source (59.5~keV gamma, activity 6~kBq) and a $^{226}\mathrm{Ra}$ source (186~keV gamma, activity $960$~Bq) are chosen from the set of calibration sources, as they cover the relevant low-energy range of interest for the MM trigger.

\begin{figure}[htbp]
    \centering
    \begin{subfigure}{0.48\textwidth}
        \centering
        \includegraphics[width=\textwidth]{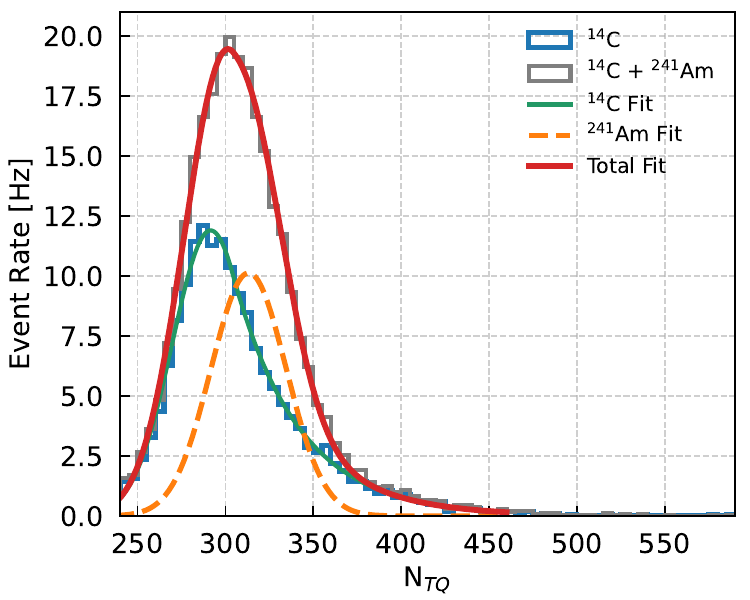}
        \caption{}
    \end{subfigure}%
    \hfill
    \begin{subfigure}{0.48\textwidth}
        \centering
        \includegraphics[width=\textwidth]{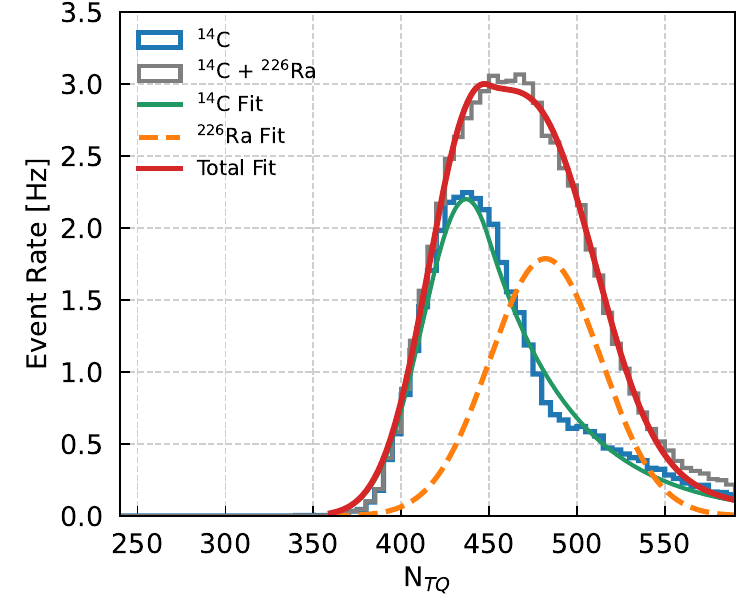}
        \caption{}
    \end{subfigure}
    \caption{Simulation (a): The number of T/Q pairs ($N_{TQ}$) spectrum of pure $^{14}$C and $^{14}$C + $^{241}$Am after reconstruction cut and 25 keV energy threshold. (b): The number of T/Q pairs ($N_{TQ}$) spectrum of pure $^{14}$C and $^{14}$C + $^{226}$Ra after reconstruction cut and 110 keV energy threshold. In both simulations, the calibration sources are positioned at the center of the detector ($z = 0$).}
    \label{fig:calib}
\end{figure}

We performed simulations combining $^{14}\mathrm{C}$ with either $^{241}\mathrm{Am}$ or $^{226}\mathrm{Ra}$ and applied three visible-energy selection thresholds: 25, 72 and 110 keV. 
$N_{TQ}$ denotes the number of T/Q pairs associated with each MM-triggered event within a 700~ns time window.
The raw $N_{TQ}$ distributions from the calibration sources are difficult to separate from the $^{14}\mathrm{C}$ background due to the high intrinsic $^{14}\mathrm{C}$ rate.
A reconstruction algorithm (described in Appendix~\ref{appendix:reco}) is therefore applied; events are selected according to $|r_{\mathrm{reco}}-r_{\mathrm{source}}| < r_{\mathrm{cut}}$. 
Two examples of $N_{TQ}$ spectra after this spatial selection are shown in Figure~\ref{fig:calib}. 
The choice of $r_{\mathrm{cut}}$ is optimized to balance signal statistics and background rejection.
For the low energy $^{241}\mathrm{Am}$ source ($59.5$~keV), a tight cut of $r_{\mathrm{cut}}=0.7$~m is adopted to maximize the signal-to-background ratio against the intense $^{14}$C background. 
Conversely, for the higher energy $^{226}\mathrm{Ra}$ source ($186$~keV), a looser cut of $r_{\mathrm{cut}}=2$~m is used to increase the statistics of calibration events.
As an example, we apply a 25~keV visible-energy threshold in Figure~\ref{fig:calib}(a) to demonstrate the trigger performance in an ideal low-threshold scenario, and a 110~keV threshold in Figure~\ref{fig:calib}(b) corresponding to the current operating conditions. 
% Meanwhile, a pure $^{14}\mathrm{C}$ spectrum is used as the control group.
Meanwhile, a pure $^{14}\mathrm{C}$ spectrum is used as the control sample.

To extract the calibration-source signal, the pure $^{14}\mathrm{C}$ spectrum is modelled with a Gaussian plus an exponential tail to account for the pile-up of multiple $^{14}\mathrm{C}$ events; the shape parameters obtained from this background fit are then fixed.
The combined spectrum ($^{14}\mathrm{C} + ${source}) is fitted by adding an additional Gaussian component representing the calibration source, and the amplitude of this component yields the measured source event rate. To account for the position-dependent response, this measured rate is averaged over six equal-volume positions along the $z$-axis. After correcting this averaged rate for the relevant factors — decay branching ratio, gamma escape probability, and reconstruction efficiency — the MM trigger efficiency is derived and presented in Figure~\ref{fig:trigger_eff_new}. 
Note that, owing to scintillator nonlinearity, the 59.5~keV gamma from $^{241}\mathrm{Am}$ corresponds to an average visible energy of approximately 39~keV, while for $^{226}\mathrm{Ra}$ the visible energy is about 135~keV. 
The horizontal error bars arise from the uncertainty in the energy deposition process, while the vertical error bars reflect the uncertainty associated with the event reconstruction procedure. 
At the 25~keV threshold, the extracted $^{241}\mathrm{Am}$ signal yields a trigger efficiency of about 92\%, in good agreement with the expectation from the efficiency curve. 
For the $^{226}\mathrm{Ra}$ source, the extracted efficiencies are approximately 95\% and 81\% at the 72~keV and 110~keV thresholds, respectively, which are also consistent with the corresponding curves. 
These results demonstrate that using radioactive calibration sources is a suitable and reliable method to validate the trigger performance with real data.

\subsection{Astrophysical Monitor}
\label{sec:astrophysical_monitor}

Astrophysical neutrinos are often emitted by transient sources, such as core-collapse supernovae, type Ia supernovae, neutron star mergers, and others. 
One of the main scientific goals of the MM trigger is to search for transient bursts of astrophysical neutrinos, with core-collapse supernovae serving as a representative benchmark (see Section~\ref{sec:ccsn}). 

To achieve this, the \textbf{Transient Neutrino Burst Monitor}, which identifies neutrino burst candidates based on real-time trigger information,
and the \textbf{External Alert Processor}, which ingests alerts from external observatories and coordinates follow-up data acquisition,
are implemented and executed in real time within the PU.

\subsubsection{Transient Neutrino Burst Monitor}
\label{sec:Supernova_Monitor}

\begin{figure}[htbp]
\begin{minipage}{6in}
    \centering
{\includegraphics[width=0.99\columnwidth]{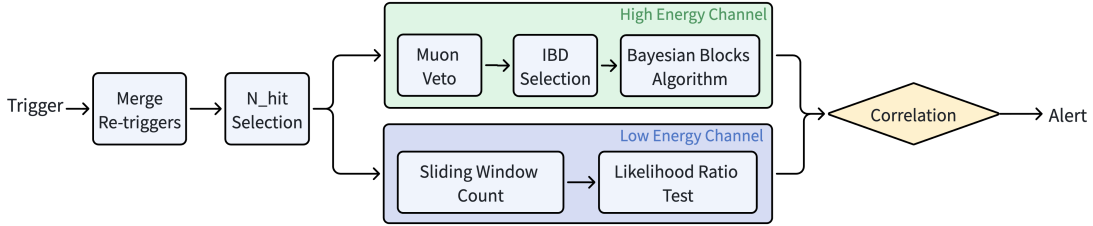}}
\end{minipage}
   \centering
  \caption{The workflow of Transient Neutrino Burst Monitor.}
  \label{fig:monitor_strategy}
\end{figure}

The workflow of the Transient Neutrino Burst Monitor is demonstrated using data collected in 2025, as illustrated in Figure~\ref{fig:monitor_strategy},
while the corresponding parameter configuration is summarized in Table~\ref{table:configure}.
Each trigger contains only two key quantities: the trigger time and the number of trigger hits (trigger $N_{\text{hit}}$) within the trigger time window. This minimal information is intentionally used to enable a fast and low-latency alert generation, avoiding computationally expensive reconstruction steps such as position or energy reconstruction. As the trigger algorithm described in Section \ref{sec:Trigger_Algorithms} operates with a fixed time window, the first step of the workflow is to merge re-triggers originating from the same physical signal. After applying the trigger $N_{\text{hit}}$ selection, the triggers are divided into two channels: a high energy channel and a low energy channel. The high energy channel primarily targets IBD events, while the low energy channel focuses on $\nu$pES and $\nu$eES events.

\begin{table}[htbp]
\setlength{\tabcolsep}{8pt}
\renewcommand{\arraystretch}{1.3}
\centering

\vspace{0.5em}

\begin{tabular}{lclc}
\hline
\multicolumn{2}{c}{\textbf{High Energy Channel}} & \multicolumn{2}{c}{\textbf{Low Energy Channel}} \\
\hline
Parameter & Value & Parameter & Value \\
\hline
Trigger $N_{hit}$ Range & [7000, 53000] & Maximum $N_{hit}$  & 7000 \\
(Energy Range~[MeV]) & [5, 40] & (Energy Range~[MeV]) & [0.1, 5] \\
WP Veto PMT Cut & 400 & \multirow{2}{*}{Window Length~[ms]} & \multirow{2}{*}{5}\\
WP Veto Time~[ms] & [-0.01, 1.5] & & \\
IBD Delay Trigger $N_{hit}$ Range & [2685, 4285] &  \multirow{2}{*}{Baseline} & 32--37 \\
IBD Delay Time Range~[ms] & [0.01, 1] & & (updated every 10 mins) \\
\hline
\multicolumn{4}{c}{\textbf{Common Parameters}} \\
\hline
\multicolumn{2}{l}{$\mathrm{TS}_{\mathrm{total}}$ (FAR=1/week)} & \multicolumn{2}{c}{35} \\
\multicolumn{2}{l}{$\mathrm{TS}_{\mathrm{total}}$ (FAR=1/month)} & \multicolumn{2}{c}{42} \\
\hline
\end{tabular}
\caption{High and low energy channels parameter configuration. The lower boundary of the low energy channel is determined by the system trigger threshold ($-\ln\mathcal{L}$ threshold of 218), which corresponds to the lower limit of the energy range discussed in Section~\ref{sec:Trigger_Algorithms} and Section~\ref{sec:Trigger_Efficiency}.}
\vspace{0.5em}
\label{table:configure}
\end{table}

\begin{table}[htbp]
\setlength{\tabcolsep}{10pt} % 默认是6pt，改成10pt让列更宽松
\renewcommand{\arraystretch}{1.3} % 默认是1，1.3表示行距加30%
\centering
\vspace{0.5em} % 标题与表格之间留一点空
\begin{tabular}{lccc}
\hline
Energy & $2.2$ MeV & $4.4$ MeV & $6.1$ MeV \\
\hline
% Trigger $N_{hit}$ per event  & $3485 \pm 269$ & $6681 \pm 510$ & $8575 \pm 662$ \\
% Trigger $N_{hit}$ per event  & $3485 $ & $6681 $ & $8575 $ \\
Trigger $N_{hit}$ per event  & $3485 \pm 291$ & $6681 \pm 624$ & $8575 \pm 686$ \\
\hline
\end{tabular}
\caption{Average trigger $N_{hit}$ from the AmC source. The quoted standard deviations are obtained by combining in quadrature the intrinsic peak width of the $N_{hit}$ distribution and the variation of the peak position with source deployment location.}
\vspace{0.5em} % 表格和标签之间留一点空
\label{table:calibration_monitor}
\end{table}

\begin{figure}[htbp]
    \centering
    \begin{subfigure}{0.5\textwidth}
        \centering
        \includegraphics[width=\textwidth]{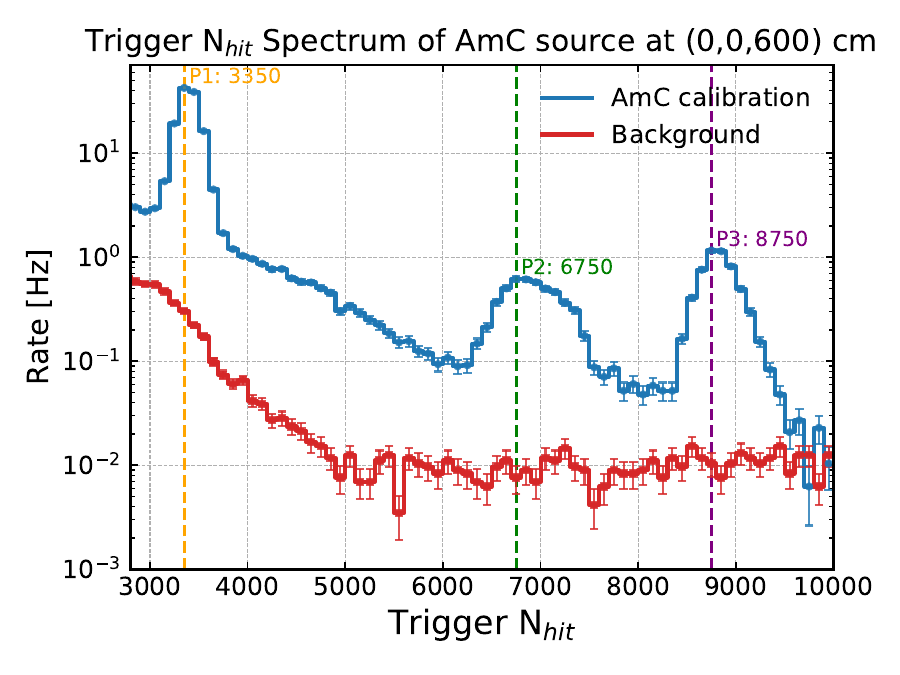}
        \caption{}
    \end{subfigure}%
    \hfill
    \begin{subfigure}{0.49\textwidth}
        \centering
        \includegraphics[width=\textwidth]{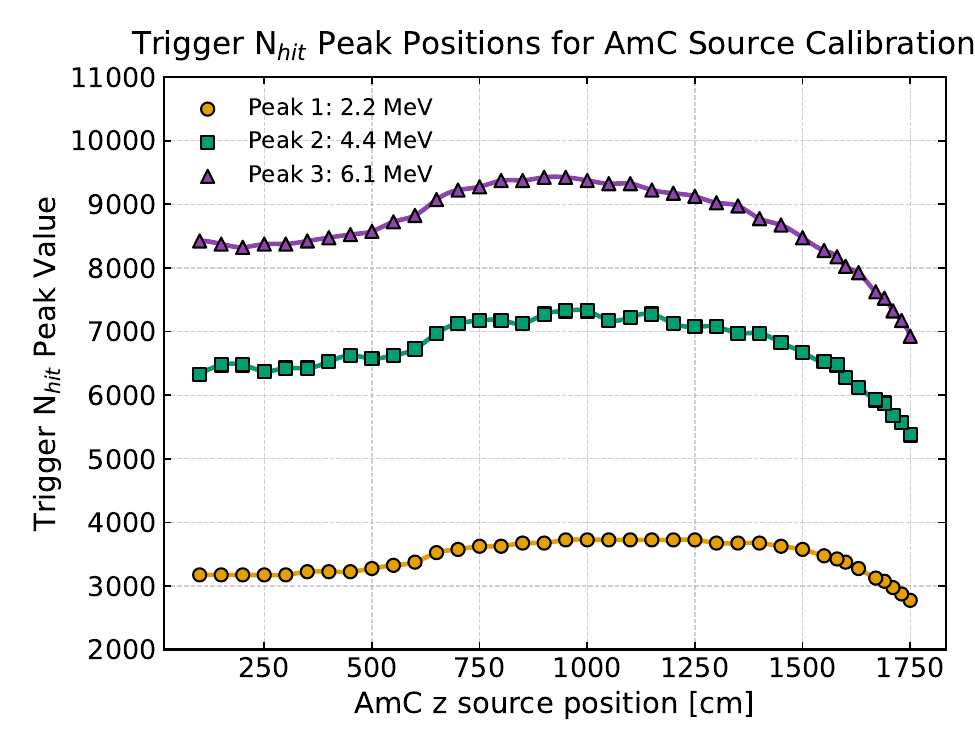}
        \caption{}
    \end{subfigure}
\caption{
(a) The trigger $N_{\mathrm{hit}}$ spectrum of the AmC source deployed at the detector position $(0, 0, 600)$~cm (blue solid line), based on data collected in 2025. 
The red solid line represents the intrinsic detector background, measured during an independent control run without the calibration source deployed, covering an equivalent time interval. 
The vertical dashed lines mark the positions of the three distinct peaks (P1, P2, and P3) corresponding to the gamma emissions from the AmC source.
(b): The trigger $N_{hit}$ peaks of three energies with different AmC source positions.
}
    \label{fig:AmC}
\end{figure}

\paragraph{Trigger N\textsubscript{hit} Calibration.}
To establish the relationship between trigger $N_{\text{hit}}$ and energy, Americium–Carbon (AmC) calibration sources were deployed in the JUNO CD\cite{Abusleme2021Calibration}. The AmC source produces ${}^{16}\mathrm{O}$ through the reaction $^{13}\mathrm{C}(\alpha, n){}^{16}\mathrm{O}$. In this process, three gamma rays with energies of 6.1~MeV, 4.4~MeV, and 2.2~MeV are emitted, and the corresponding peaks can be observed in Figure~\ref{fig:AmC}(a). Figure~\ref{fig:AmC}(b) shows the trigger $N_{\text{hit}}$ peaks obtained from a position scan along the z-axis. The $N_{\mathrm{hit}}$ value for each peak is determined directly by identifying the position of the maximum bin in the distribution. To properly account for the detector geometry, a volume-weighted average is applied, where each measurement point is weighted by the volume of its corresponding spherical shell. The resulting mean trigger $N_{\mathrm{hit}}$ values for the three energies are summarized in Table \ref{table:calibration_monitor}. 
Based on these calibration points, the boundaries of the low energy and high energy channels are determined through linear interpolation and extrapolation. The trigger $N_{\mathrm{hit}}$ cut of the high energy channel is set to [7000, 53000], which corresponds approximately to an energy range of [5, 40]~MeV. 
Similarly, the low energy channel is defined with a maximum trigger $N_{\mathrm{hit}}$ of 7000, corresponding to an upper energy limit of approximately 5~MeV. Its lower boundary is not defined by a sharp $N_{\mathrm{hit}}$ cut, but is instead intrinsically determined by the system trigger threshold ($-\ln\mathcal{L}$ threshold of 218), which corresponds to an energy of approximately 0.1~MeV.

\begin{figure}[htbp]
\begin{minipage}{6in}
    \centering
{\includegraphics[width=0.75\columnwidth]{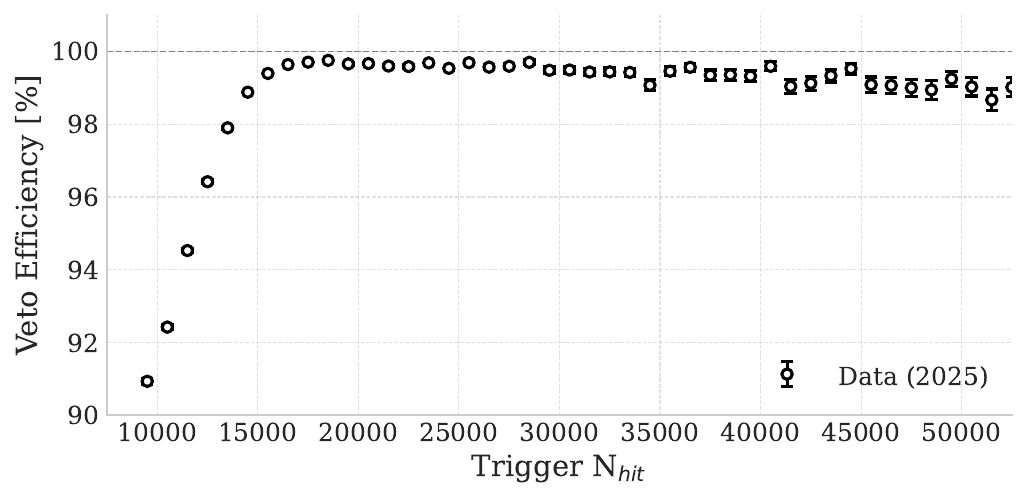}}
\end{minipage}
   \centering
  \caption{Measured muon veto efficiencies in the high energy channel. The veto condition is defined by a WP trigger threshold of $N_{\mathrm{hit}}^{\mathrm{WP}} > 400$ and a veto time window of $[-0.01, 1.5]$~ms.}
  \label{fig:muon_veto}
\end{figure}

\paragraph{Muon Veto.} 
Cosmic-ray muons constitute a dominant background in the monitoring of transient neutrino bursts. In particular, muon-induced spallation can produce bursts of correlated events that mimic genuine astrophysical signals. If not properly suppressed, such events can significantly increase the false alert rate. Therefore, using the WP PMTs to identify muons and apply a veto to CD trigger events is essential. As shown in Figure~\ref{fig:hardware_structure}, once the CTU detects more than 400 hits in the WP, it issues a WP trigger to the PU, which is used as a muon veto signal. At this threshold, the rate of cosmic-ray muons in the WP is approximately 9~Hz. 
Based on this veto signal, a veto time window of $[-0.01,\,1.5]$~ms is applied. The upper bound of 1.5~ms is chosen to suppress muon-induced secondary signals, in particular fast neutrons produced by muon interactions, while the small negative offset of $-0.01$~ms accounts for possible timing offsets between the CD and the WP, providing a safety margin for synchronization. Given the muon rate and the veto window length, this corresponds to a dead time of only about 1--2\% of the total live time. This level of dead time is considered acceptable given the effective suppression of muon-induced backgrounds. After applying the muon veto, backgrounds from muons and fast neutrons --- with typical lifetimes of around 200~\(\mu\)s --- can be efficiently rejected. Long-lived daughter isotopes such as \(^9\)Li and \(^8\)He remain as residual backgrounds. Figure~\ref{fig:muon_veto} shows the measured veto efficiency in the high energy channel. 
The efficiency is calculated in a data-driven way as the number of CD high-energy trigger events ($N_{\text{hit}} \in [7000, 53000]$, dominantly muons) that have a coincident WP trigger, divided by the total number of such CD high-energy trigger events. The average efficiency is approximately 97\%. However, this performance gradually improves with increasing $N_{\text{hit}}$, and the veto efficiency can reach over 99\% in the range of $[20000, 50000]$.

\paragraph{IBD Selection.} 
The IBD channel provides a clean signature for $\bar{\nu}_e$ detection in JUNO, with a characteristic prompt--delayed coincidence topology that enables efficient background suppression. The prompt signal originates from a positron, followed by annihilation into two 0.511~MeV $\gamma$-rays. The delayed signal comes from a 2.2~MeV $\gamma$-ray emitted when the neutron is captured by a proton, approximately 200~$\mu$s after the prompt signal. In the Transient Neutrino Burst Monitor, only the energy and time-coincidence information is used, while spatial reconstruction is intentionally omitted to ensure a fast and low-latency monitoring process. 
Based on the AmC calibration runs, the time window and trigger $N_{\mathrm{hit}}$ intervals for the delayed signal are chosen to be [10~\textmu s, 1~ms] and [2685, 4285], respectively. 
The lower bound of 10~\textmu s is chosen to suppress PMT after pulses, while the upper bound of 1~ms corresponds to approximately five times the neutron capture time, ensuring high neutron detection efficiency.
The $N_{\mathrm{hit}}$ interval is determined based on the 2.2~MeV peak at 3485 from Table~\ref{table:calibration_monitor}, which is extended to $3485 \pm 800$. This range is applied to both the AmC and background data to achieve the best signal-to-background ratio shown in Figure~\ref{fig:AmC}(a). With this delayed-signal selection applied to the AmC source data, the neutron selection efficiency is about 92\%.

\begin{figure}[htbp]
    \centering
    \begin{subfigure}{0.45\textwidth}
        \centering
        \includegraphics[width=\textwidth]{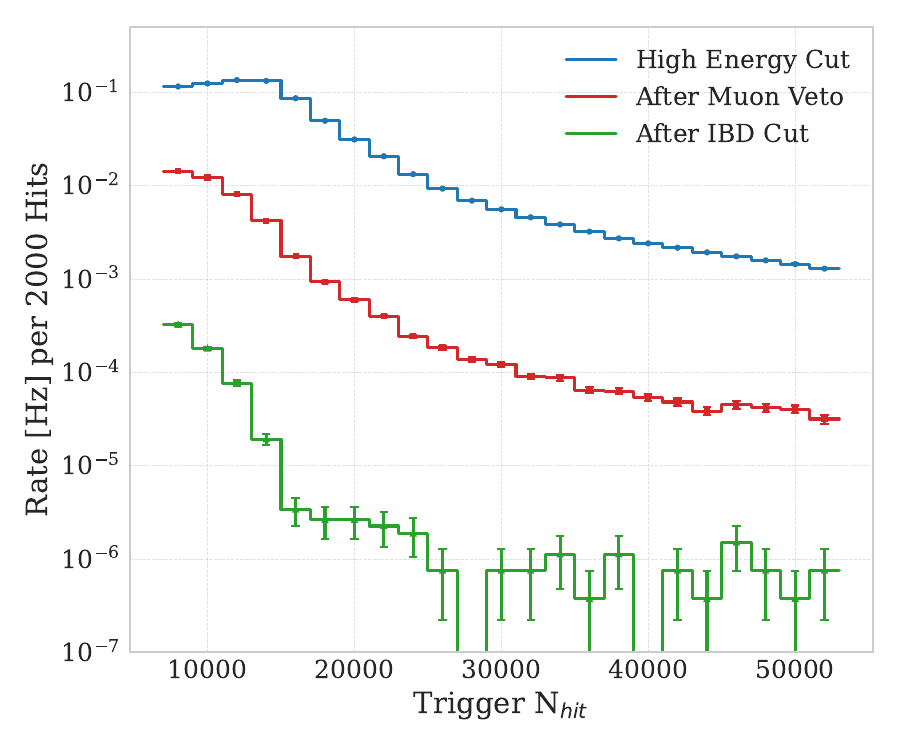}
        \caption{}
        % \caption{The background data of high energy channel.}
    \end{subfigure}%
    \hfill
    \begin{subfigure}{0.48\textwidth}
        \centering
        \includegraphics[width=\textwidth]{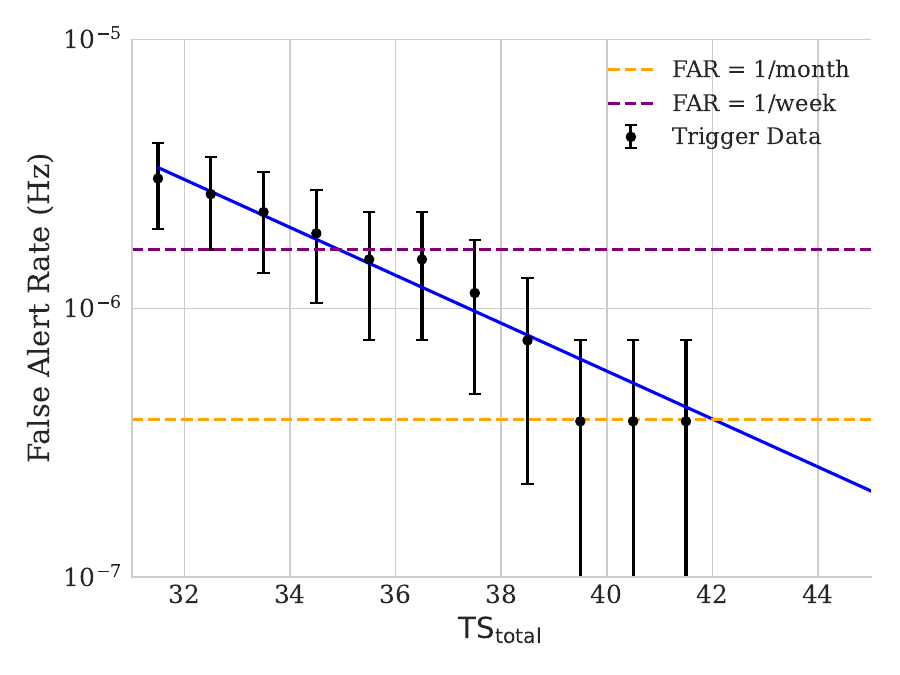}
        \caption{}
        % \caption{The FAR distribution with $\mathrm{TS}_{\mathrm{total}}$.}
    \end{subfigure}
    \caption{(a) The background data of high energy channel. (b) The FAR versus $\mathrm{TS}_{\mathrm{total}}$.}
    \label{fig:IBD_bg}
\end{figure}

We summarize the impact of the successive background reduction steps in the high energy channel as follows. As shown in Figure~\ref{fig:IBD_bg}(a), the application of the trigger $N_{\mathrm{hit}}$ selection results in a background rate of approximately 0.8~Hz, dominated by cosmic-ray muons. The subsequent muon veto suppresses this rate to about 0.03~Hz. With the addition of the IBD delayed coincidence selection, the background is further reduced to $\sim$64 events per day (0.00074~Hz), achieving a sufficiently low level for reliable burst monitoring in the high energy channel. The residual background is primarily composed of reactor neutrinos, remaining muons, and muon-induced correlated events.

\paragraph{Bayesian Blocks Algorithm.} 
After applying IBD selection in the high energy channel, the corresponding trigger is regarded as a high energy event candidate, and its event time is used as input to the monitoring algorithm. 
The monitoring algorithm employed is the Bayesian Blocks Algorithm (BBA), which is a model-independent method that makes no assumptions regarding the time profile, time scale, or signal strength~\cite{Scargle1998, Scargle2013}. The BBA aims to determine whether the event rate within a given time interval is statistically stable. This is achieved by fitting the event rate in real time using a piece-wise constant model. Technically, the time interval is divided into one or more segments, each referred to as a \emph{block}. Within each block, the event rate is assumed to be constant, but it may change at the boundaries between blocks. The best-fit model is obtained by maximizing a likelihood defined using the Cash statistic~\cite{Cash1979}. In practice, the BBA is applied to a time-ordered sequence of the most recent 50 high energy event candidates. For further technical details on the BBA implementation in the MM trigger, refer to Appendix~\ref{appendix:BBA} and the original references~\cite{Scargle1998, Scargle2013}. The BBA produces a test statistic (TS) to quantify the presence of a burst:

\[
\mathrm{TS}_{\text{high}} =
\begin{cases}
-2 \ln \left( \dfrac{L_{0}}{L_{1}} \right), & \text{if } L_{1} > L_{0} \\
0, & \text{otherwise}
\end{cases}
\]

Here, \(L_0\) denotes the likelihood under the single-block hypothesis, and \(L_1\) corresponds to the likelihood under a multi-block hypothesis with an increasing block rate.

\paragraph{Low Energy Channel.}

In the low energy channel, the monitoring is performed by counting the number of triggers within a fixed 5~ms time window and comparing it with an expected baseline. The choice of the 5~ms window is motivated by the fact that low energy neutrino signals are expected to precede the IBD-dominated high energy signal by $\sim$10~ms, as will be discussed in Section~\ref{sec:low_energy_channel}.

\[
\mathrm{TS}_{\text{low}} =
\begin{cases}
2 \left[ n \ln\left( \dfrac{n}{b} \right) - (n - b) \right], & \text{if } n > b \\
0, & \text{if } n \leq b
\end{cases}
\]

Here, \(n\) is the number of low energy triggers observed within the 5~ms window, and \(b\) is the expected number of triggers estimated from a baseline. The baseline is dynamically evaluated every 10 minutes using the average trigger rate in the low energy channel, and typically falls within the range of 32--37 counts per 5~ms window, as summarized in Table~\ref{table:configure}.

\paragraph{Correlation and Alert Threshold.}

The final step in the workflow of the Transient Neutrino Burst Monitor is to correlate the high energy and low energy channels in order to form the alert decision. In the current implementation, an alert condition is first identified in the high energy channel using the BBA, requiring the detection of a multi-block structure with an increasing block rate. Only if a high energy burst is identified, the corresponding test statistic $\mathrm{TS}_{\text{high}}$ is combined with the low energy test statistic $\mathrm{TS}_{\text{low}}$, which is evaluated within the 5~ms time window associated with the last high energy event candidate. The total test statistic is then defined as

\[
\mathrm{TS}_{\mathrm{total}} = \mathrm{TS}_{\text{low}} + \mathrm{TS}_{\text{high}}
\]

In typical background events, $\mathrm{TS}_{\mathrm{total}}$ is predominantly driven by $\mathrm{TS}_{\text{high}}$, with $\mathrm{TS}_{\text{low}}$ providing only a minor contribution. To determine the alert threshold, the relationship between $\mathrm{TS}_{\mathrm{total}}$ and the false alert rate (FAR) is evaluated using one month of 2025 data through an offline scan, as shown in Figure~\ref{fig:IBD_bg}(b). Based on this study, the threshold values of $\mathrm{TS}_{\mathrm{total}}=35$ and $42$ correspond to FARs of 1/week and 1/month, respectively.

\paragraph{A False Alert Example.}

\begin{figure}[htbp]
\begin{minipage}{6in}
    \centering
{\includegraphics[width=0.99\columnwidth]{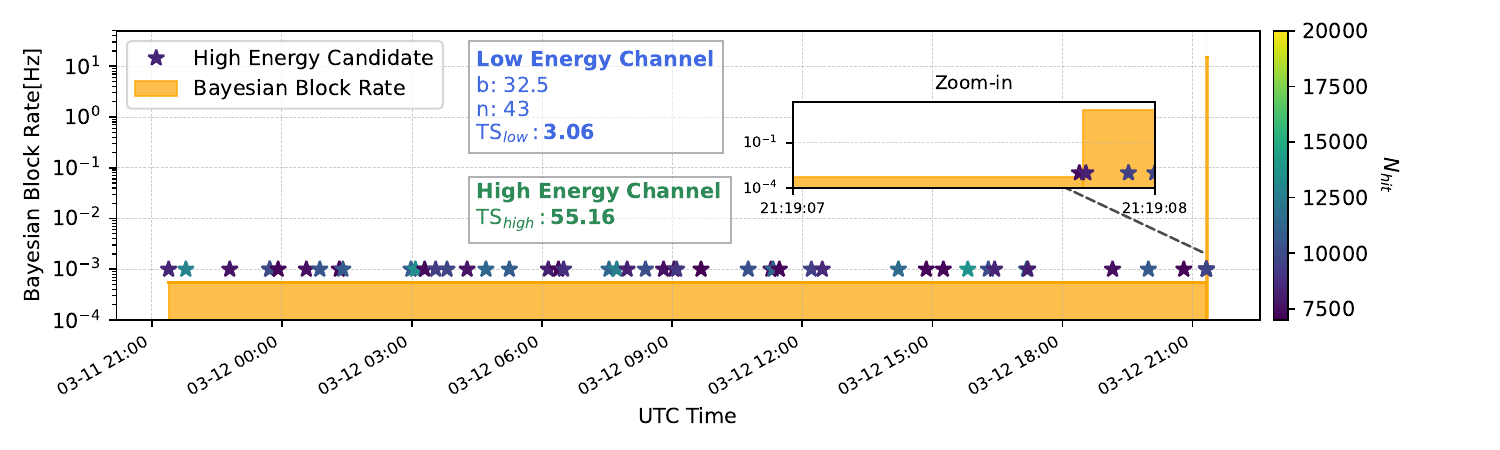}}
\end{minipage}
   \centering
    \caption{A false alert example observed in March 2026. The star markers denote high energy event candidates, and the shaded region shows the Bayesian Block rate. The inset shows a zoomed-in view of the candidate cluster associated with the alert, in the 5~ms window used for the low energy channel. The annotated values in the main panel indicate the resulting $\mathrm{TS}_{\text{high}}$ and $\mathrm{TS}_{\text{low}}$ for this event.}
  \label{fig:false_alert}
\end{figure}

An example of a false alert observed in March 2026 is shown in Figure~\ref{fig:false_alert}. 
% In this case, the alert is issued by a high energy muon. Approximately 0.1~s before the alert time, a muon event with $N_{\mathrm{hit}} > 5\times10^{6}$ is observed. 
In this case, the alert is issued by a high energy muon. Approximately 0.1~s before the alert time, a muon event with $N_{\mathrm{hit}} > 5\times10^{6}$ is observed, whose duration significantly exceeds the 1.5~ms online muon veto window.
This event is followed by four high energy events occurring within a short time interval, as shown in the inset. The BBA partitions the corresponding time series into two blocks with an increasing event rate, satisfying the burst condition in the high energy channel. This results in a high test statistic of $\mathrm{TS}_{\text{high}} = 55.16$. Within the associated 5~ms window, the low energy channel yields $\mathrm{TS}_{\text{low}} = 3.06$. The combined test statistic therefore exceeds the predefined alert threshold, issuing the alert. While such cases are very rare, future strategies will apply different veto window lengths based on the muon energy. This example illustrates that the current Transient Neutrino Burst Monitor is primarily driven by the high energy channel. The correlation procedure is initiated by the BBA condition in the high energy channel, and the total test statistic is typically dominated by $\mathrm{TS}_{\text{high}}$. Future developments of the monitor will focus on enhancing the role of the low energy channel.

\subsubsection{DAQ Response to Neutrino Burst Alerts}
\label{sec:DAQ_Response_MM}

Once a transient neutrino burst is identified by the Transient Neutrino Burst Monitor, the MM trigger sends an alert to the DAQ system, which then switches to a dedicated trigger-less acquisition mode. In this mode, the full charge and time information of all PMT signals are continuously recorded, ensuring that no potentially relevant data is lost.

Given the extreme rarity of astrophysical neutrino bursts and the low false alert rate achieved by the monitor, this trigger-less mode is implemented to maximize the completeness of the recorded dataset without causing excessive data throughput. For a typical core-collapse supernova neutrino burst, which lasts on the order of 10~s, the DAQ system maintains this mode for a time window of $\pm$60~s around the alert time. This strategy guarantees that the full temporal evolution of the neutrino signal is captured, including possible early or delayed components beyond the main burst duration, thereby providing a comprehensive dataset for subsequent offline analysis.

\subsubsection{External Alert Processor}
\label{sec:External_Alert_Processor}

\begin{figure}[htbp]
\begin{minipage}{6in}
    \centering
    \includegraphics[width=0.60\columnwidth]{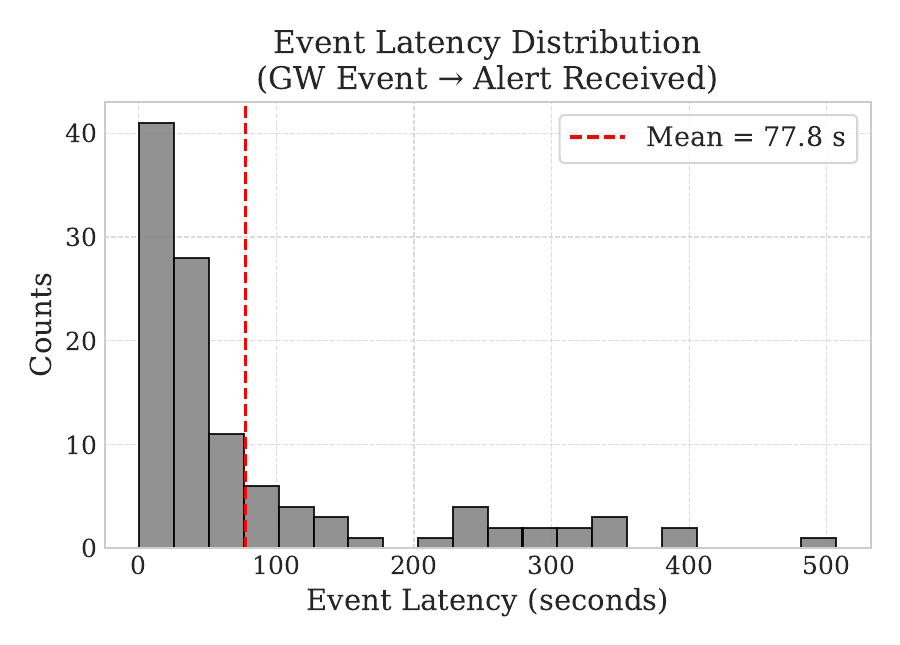}
\end{minipage}
\centering
\caption{Distribution of the latency for LVK gravitational-wave alerts. This latency is defined as the total time difference between the actual gravitational-wave event and the moment the alert is received by the PU.}
\label{fig:gw_alert_latency}
\end{figure}

The External Alert Processor continuously receives alert notifications from multiple external experiments via the General Coordinates Network (GCN) webpage~\cite{GCNweb}, including signals from gravitational-wave detectors, gamma-ray satellites, and other transient astrophysical observatories. For example, between June 18 and July 18, 2025, 111 gravitational-wave alerts~\cite{Chaudhary2024LowLatencyGW} were received by the PU. The latency between the gravitational-wave event time and the alert receipt time is shown in Figure~\ref{fig:gw_alert_latency}. The mean latency is about 77 seconds, with a maximum recorded latency of approximately 500 seconds. Upon receiving an external alert, the PU sends a tag to the DAQ system, which enables an extended data acquisition mode. In this mode, instead of the nominal $\sim$1\% data transfer, all MM trigger event data within a time window of $\pm$20~minutes around the alert time are recorded and transferred to the IHEP data centre for analysis (see Section~\ref{sec:DAQ}). This $\pm$20-minute window is chosen based on the maximum alert latency. It safely covers the 500 seconds delay, ensuring that the complete MM trigger event data around the true astrophysical event time is captured.

Such alerts are particularly important for events like binary neutron star mergers or neutron star–black hole mergers, which are expected to emit not only gravitational waves but also MeV‑scale neutrinos and potentially gamma‑ray bursts. Coordinated multi‑messenger observations of these events allow for coincidence analyses between gravitational‑wave signals and neutrino detections, significantly enhancing sensitivity to weak neutrino signals while reducing background. This, in turn, enables detailed studies of the astrophysical source properties, including the total energy release, neutrino emission mechanisms, and jet dynamics\cite{Sekiguchi2011, Abbott2017_GW170817,Cusinato2022BNS}.

\section{System Performance for Supernova Neutrino Bursts}
\label{sec:ccsn}
Massive stars with $M \geq 8\,M_\odot$ end their lives in core collapse supernovae (CCSN)~\cite{Smartt2009}. The gravitational collapse of the stellar core releases a gigantic amount of energy in a very short time. 
The gravitational collapse produces a very hot and dense core, in which neutrinos are created in huge numbers. Since only neutrinos can escape from this dense environment, they serve as ideal messengers for observing CCSNe. Neutrino emission from a CCSN proceeds through several distinct phases: core collapse and neutrino trapping, neutronization burst, accretion, and cooling~\cite{Thompson2003}.
During the accretion phase, neutrinos are emitted from the hot, dense environment surrounding the proto-neutron star (PNS), offering direct insight into the explosion mechanism~\cite{sukhbold2016core}. Consequently, precisely reconstructing the neutrino light curve (time evolution) across these early phases is crucial for unraveling the dynamics of the supernova explosion.

\begin{figure}[htbp]
  \centering
  \includegraphics[width=0.55\textwidth]{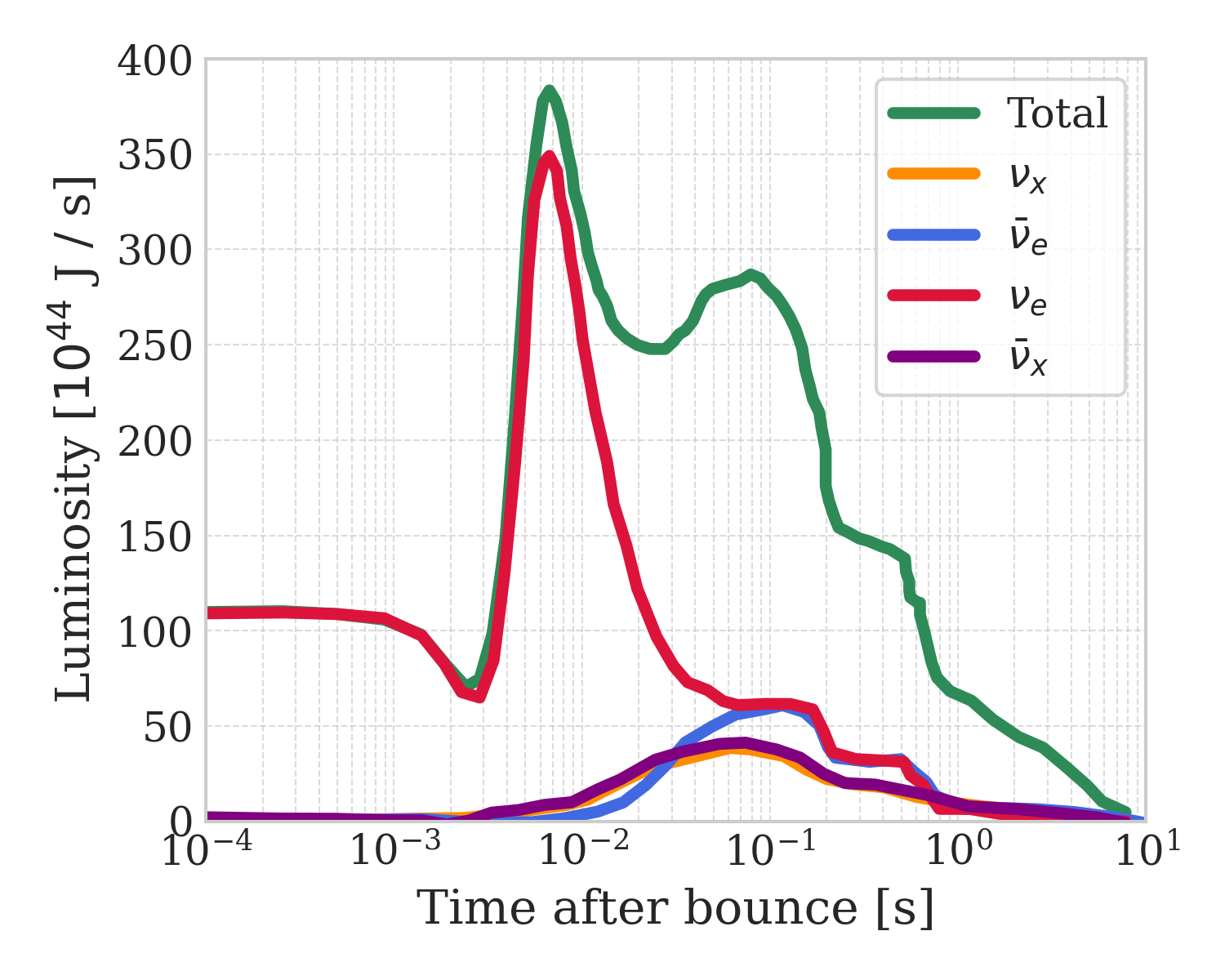}
  \caption{Neutrino luminosity of a 27 $M_\odot$ core-collapse supernova (Garching model~\cite{Mirizzi2016}). Here, $\nu_x$ represents the heavy-flavor neutrinos ($\nu_\mu$ and $\nu_\tau$), and $\bar{\nu}_x$ represents their corresponding antineutrinos.}
  \label{fig:sn_luminosity}
\end{figure}

Besides probing the astrophysical explosion mechanism, the stellar core collapse also provides a unique opportunity to study fundamental physics under extreme conditions that are not accessible in terrestrial laboratories~\cite{Mirizzi2016}. Furthermore, in the subsequent cooling phase, neutrinos of all flavors are emitted with approximately thermal spectra. This allows for detailed studies of PNS properties and enables the search for potential exotic physics, such as axions or sterile neutrinos~\cite{keil2003monte,chang2018supernova}.

Figure~\ref{fig:sn_luminosity} shows the neutrino luminosity for different neutrino flavors of a $27\,M_\odot$ CCSN as an illustrative example. The observations of neutrinos from all these phases are essential for testing CCSN models.

\subsection{Alert Performance Based on Real Background}
\label{sec:alert_performence}

Building upon the workflow and background levels established in Section~\ref{sec:Supernova_Monitor}, this section evaluates the overall alert performance of the Transient Neutrino Burst Monitor. Here, we estimate the FAR using real background data to determine the required alert thresholds and utilize simulated supernova neutrinos as the signal input to assess the alert efficiency and alert time.

In this work, four representative supernova neutrino models are adopted, including two models from the Nakazato simulations~\cite{nakazato2013} and two from the Garching simulations~\cite{Mirizzi2016}. The Nakazato models correspond to progenitor masses of 13~$M_\odot$ and 30~$M_\odot$, both with solar metallicity ($Z=0.02$) and a shock revival time of 300~ms. The Garching models correspond to progenitor masses of 11.2~$M_\odot$ and 27~$M_\odot$, both employing the Shen equation of state, with configurations that include convection and nucleon potential effects. These four models are chosen to represent a range of progenitor masses and provide a benchmark set for evaluating the detector response to core-collapse supernova neutrino bursts.

Further out from the neutrino-sphere --- the effective surface of last scattering where neutrinos decouple from dense stellar matter and begin to stream freely --- the flavor transformation of supernova burst neutrinos is governed by the Mikheyev–Smirnov–Wolfenstein (MSW) matter effects \cite{Wolfenstein1978, ms1985}. These effects modify the neutrino energy spectra depending on the matter density profile and the neutrino mass ordering. As a result, the flavor composition of neutrinos arriving at Earth is determined by the initial fluxes at the source and the MSW-induced transition probabilities.

\begin{figure}[htbp]
\begin{minipage}{6in}
    \centering
{\includegraphics[width=0.95\columnwidth]{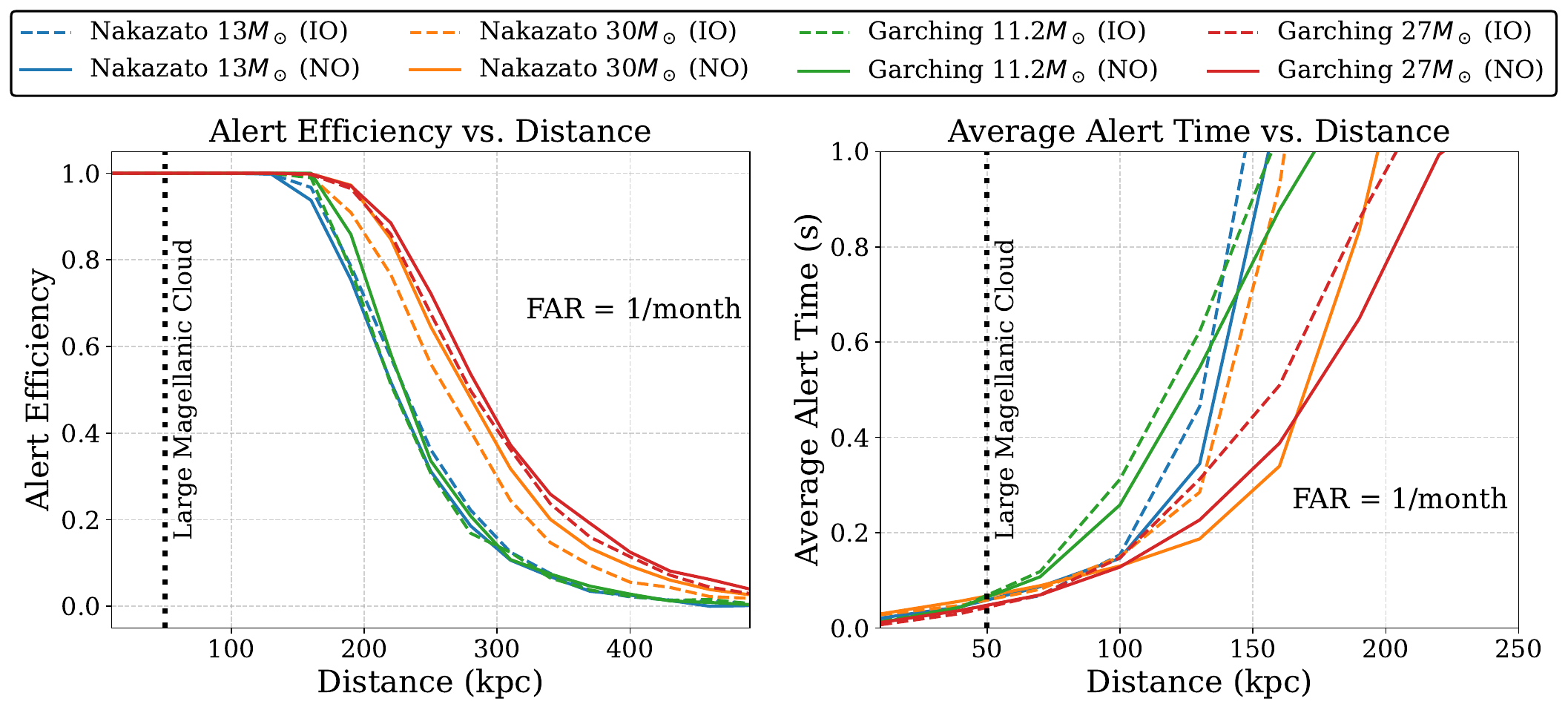}}
\end{minipage}
   \centering
  \caption{
    Expected alert efficiency (left panel) and alert time relative to the core-bounce moment (right panel) for the Transient Neutrino Burst Monitor with supernova neutrinos from Nakazato (13 and 30~$M_\odot$) and Garching (11.2 and 27~$M_\odot$) models at different distances. Both normal ordering (NO) and inverted ordering (IO) are shown. The false alert rate (FAR) is 1/month.
    }
  \label{fig:monitor_performence}
\end{figure}

After accounting for the MSW flavor transitions within the progenitor, scaling the incident neutrino flux according to the distance of the CCSN, and applying the channel-dependent cross sections in the detector, the supernova neutrino interactions are generated. To obtain realistic hit signals, the full detector response is processed using the official JUNO simulation software~\cite{lin2023junosw}, which includes the simulation of optical photon propagation, PMT responses, and front-end electronics. As detailed in Section~\ref{sec:Supernova_Monitor}, the background is evaluated using events directly sampled from data. The simulated supernova hit signals are then mixed with the real background data and used as the input to the Transient Neutrino Burst Monitor. By applying the purely data-driven threshold parameter of $\mathrm{TS}_{\mathrm{total}} = 42$, which ensures a FAR of one per month, the resulting alert performance is evaluated and presented in Figure~\ref{fig:monitor_performence}. Under this configuration, the Transient Neutrino Burst Monitor is capable of achieving a 100\% alert efficiency for 4 CCSN models within 100~kpc, covering the Milky Way as well as the Large and Small Magellanic Clouds.
Beyond this distance, the detection efficiency gradually decreases. However, as illustrated in the left panel of Figure~\ref{fig:monitor_performence}, the monitor still maintains a 50\% alert efficiency up to distances ranging from approximately 220~kpc to 280~kpc, depending on the progenitor mass and the specific emission model. Furthermore, the right panel of Figure~\ref{fig:monitor_performence} demonstrates the alert time of the system. For a typical supernova located in the Large Magellanic Cloud, the alert can be issued within less than 0.1~seconds after the core bounce. 

Compared to the Prompt Monitor and the Online Monitor introduced in Ref.~\cite{Abusleme2024SNMonitoring}, the Transient Neutrino Burst Monitor exhibits a reduced detection reach at large distances. In particular, the distance corresponding to a 50\% alert efficiency is smaller than the $\sim$350~kpc reported in previous studies. This difference mainly arises from the use of real data to determine the trigger threshold in this work, whereas previous studies relied on Monte Carlo simulations. In real data, additional backgrounds not fully modeled in simulations, such as flasher PMTs\cite{JUNO_Initial_Performance_2025}, require a more conservative threshold setting and therefore lead to a reduced sensitivity.

Despite this reduction, the Transient Neutrino Burst Monitor offers important advantages. In particular, it employs a Bayesian Blocks-based algorithm, which is largely model-independent and therefore robust against uncertainties in the neutrino emission models. In addition, the inclusion of a dedicated low energy channel further enhances the sensitivity to soft neutrino signals. The potential of the low energy channel for independent alerts will be discussed in Section~\ref{sec:low_energy_channel}.

Together with the MM-trigger system, the three monitoring systems provide independent alert channels, ensuring a more robust and reliable supernova neutrino alert system for JUNO. In addition to performing trigger-less data acquisition and storage, the DAQ system transmits alert information from these monitoring systems to the SNEWS 2.0 network~\cite{AlKharusi2021SNEWS2}. This allows alerts from different monitoring systems to be shared and supports the issuance of combined alerts.

\subsection{Potential of the Low Energy Channel for Independent Alerts}
\label{sec:low_energy_channel}

In the current configuration of the Transient Neutrino Burst Monitor, the low energy channel is only evaluated after a burst is identified in the high energy channel. This choice is motivated by the fact that the IBD-dominated high energy channel is relatively clean, while the low energy channel is more susceptible to non-physical events, such as flasher PMTs. Nevertheless, the low energy channel has significant potential for independent alerts. To explore this possibility, we consider the performance of the low energy channel alone. Since the background level, including non-physical contributions, is not yet stably determined from real data, the alert threshold is derived from simulations. In particular, a threshold of $\mathrm{TS}_{\mathrm{low}} = 25$, corresponds to a false alert rate (FAR) of 1/month under the assumption of Poisson fluctuations of baseline.

\begin{figure}[htbp]
\begin{minipage}{6in}
    \centering
    \includegraphics[width=0.98\columnwidth]{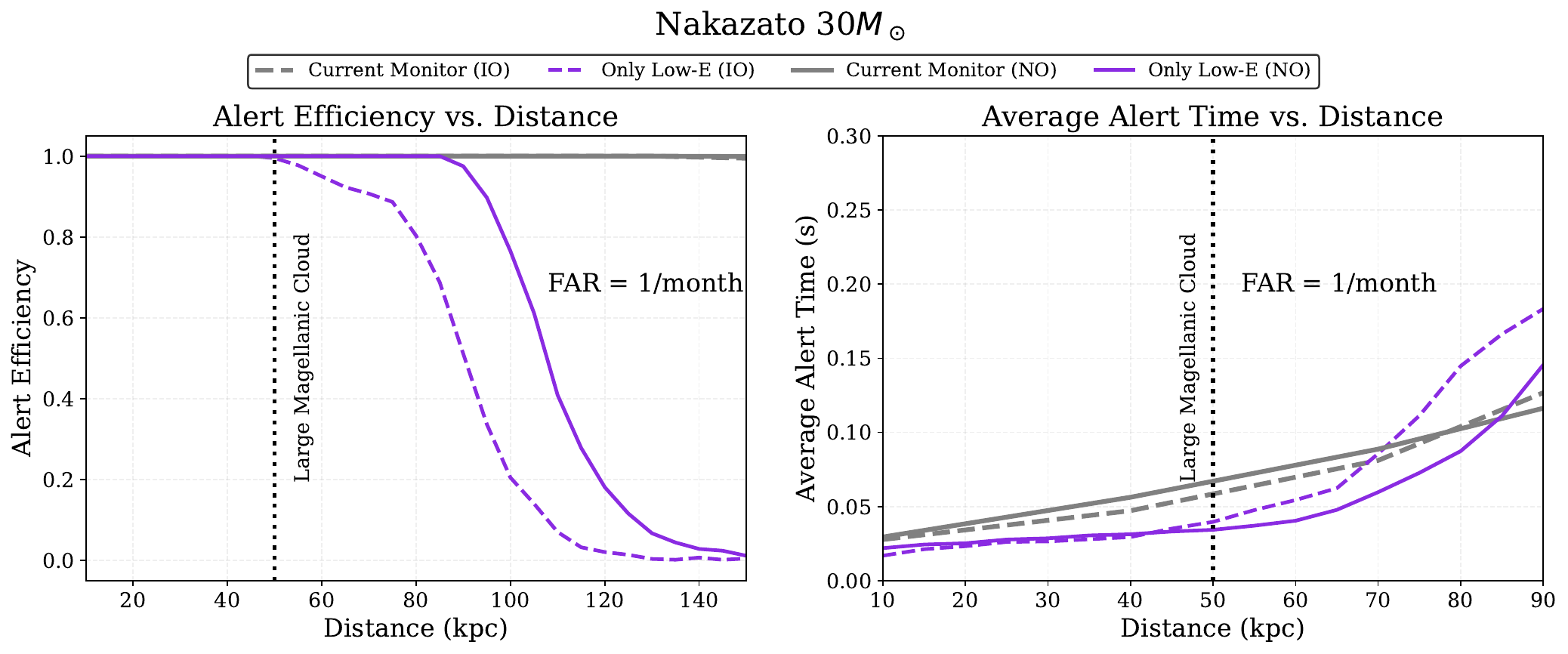}
\end{minipage}
\centering
\caption{
Alert performance of the low energy channel for supernova bursts, shown for the Nakazato 30~$M_\odot$ model. The alert efficiency (left) and the alert time relative to the core-bounce moment (right) are compared between the current Transient Neutrino Burst Monitor and the low energy channel alone, for both normal ordering (NO) and inverted ordering (IO). A false alert rate (FAR) of 1/month is assumed.
}
\label{fig:monitor_performence_low}
\end{figure}

As shown in Figure~\ref{fig:monitor_performence_low}, taking the Nakazato 30~$M_\odot$ model as an example, we compare the performance of the current Transient Neutrino Burst Monitor with that of the low energy channel alone. In terms of alert efficiency, the low energy channel is able to achieve full coverage of the Large Magellanic Cloud (LMC), demonstrating that it can independently provide effective alerts for typical Galactic and nearby extragalactic supernovae. Regarding the alert timing, at close distances (e.g., $\sim$10~kpc), the low energy channel can issue alerts earlier than the current high energy dominated monitor by approximately $\sim$10~ms. This improvement arises because the low energy channel is sensitive to all neutrino flavors through additional interaction channels such as $\nu$pES and $\nu$eES. As illustrated in Figure~\ref{fig:sn_luminosity}, the $\nu_e$ emission precedes the $\bar{\nu}_e$ emission, leading to an earlier rise of the signal in the low energy channel. As the distance increases (e.g., beyond $\sim$70~kpc), the advantage in alert timing gradually diminishes due to the reduced statistics in the low energy channel. Nevertheless, the improved promptness at close distances clearly demonstrates the potential of the low energy channel for early and independent supernova neutrino alerts.

\section{Conclusion}
\label{sec:Conclusion}
% In this work, we have presented the JUNO Multi-Messenger trigger, a low-threshold and real-time trigger system designed to extend JUNO’s sensitivity to sub-MeV neutrino signals.
In this work, we have presented the JUNO Multi-Messenger trigger, a low-threshold trigger and real-time alert system designed to extend JUNO's sensitivity to sub-MeV neutrino signals. The system achieves a stable trigger rate of approximately 7~kHz and an effective energy threshold of about $110 \pm 10$~keV, significantly lower than the global trigger. Using early data collected in 2025, we demonstrate the working principle of the Transient Neutrino Burst Monitor and validate its real-time performance. The performance of the monitor has been further evaluated using representative core-collapse supernova models. The use of real data for threshold determination results in a more conservative sensitivity compared to previous simulation-based estimates, reflecting the impact of detector backgrounds. In addition, the low energy channel has been investigated for its potential in independent alerts. The MM trigger operates in parallel with other monitoring systems in JUNO and is integrated with the DAQ to provide automated alerts to the SNEWS network. In summary, the MM trigger provides a stable and effective solution for low-threshold neutrino detection and real-time burst monitoring, enhancing JUNO’s capability in supernova neutrino observations and multi-messenger astronomy.

\section*{Acknowledgments}

We are grateful for the ongoing cooperation from the China General Nuclear Power Group. This work was supported in part by: 
the Chinese Academy of Sciences, 
the National Key R\&D Program of China, 
the Guangdong provincial government, 
and the Tsung-Dao Lee Institute of Shanghai Jiao Tong University in China, 
the Institut National de Physique Nucl\'eaire et de Physique de Particules (IN2P3) in France, 
the Istituto Nazionale di Fisica Nucleare (INFN) in Italy, 
the Fond de la Recherche Scientifique (F.R.S.-FNRS) and the Institut Interuniversitaire des Sciences Nucl\'eaires (IISN) in Belgium, 
the Conselho Nacional de Desenvolvimento Cient\'ifico e Tecnol\'ogico in Brazil, 
the Agencia Nacional de Investigaci\'on y Desarrollo and ANID - Millennium Science Initiative Program - ICN2019\_044 in Chile, 
the European Structural and Investment Funds, the Czech Ministry of Education, Youth and Sports and the Charles University Research Center in Czech Republic, 
Deutsche Forschungsgemeinschaft (DFG), the Helmholtz Association, and the Cluster of Excellence PRISMA+ in Germany, 
the Joint Institute for Nuclear Research (JINR) and Lomonosov Moscow State University in Russia, 
the Slovak Research and Development Agency in Slovak, 
MOST and MOE in Taipei, 
the Program Management Unit for Human Resources \& Institutional Development, Research and Innovation, Chulalongkorn University, and Suranaree University of Technology in Thailand, 
the Science and Technology Facilities Council (STFC) in the UK, 
and the University of California at Irvine and the National Science Foundation in the US.

\appendix

\section{Firmware} 
\label{appendix:Firmware}

\begin{figure}[htbp]
    \centering
    \begin{subfigure}{0.53\textwidth}
        \centering
        \includegraphics[width=\textwidth]{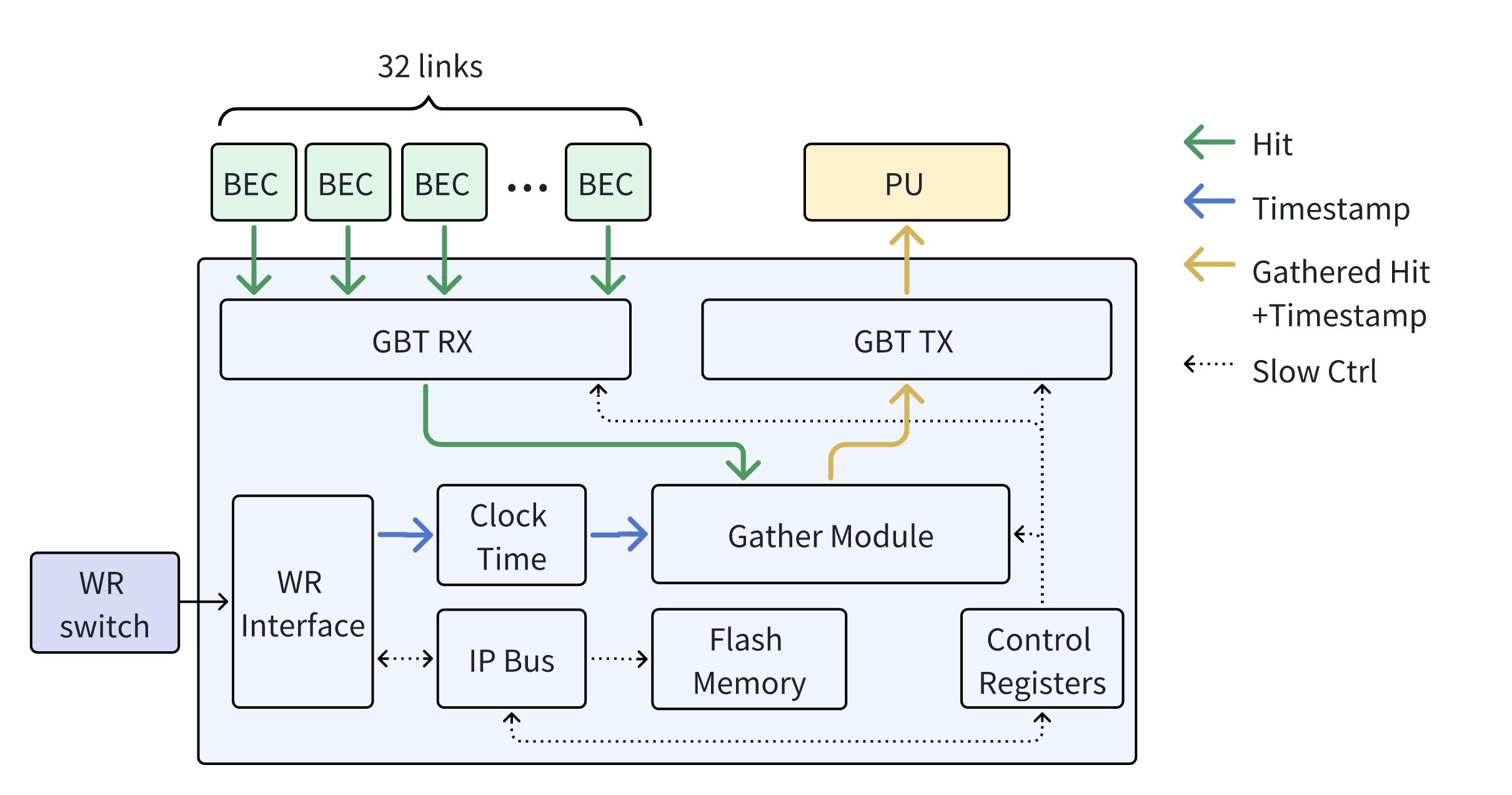}
        \caption{Illustration of the GU firmware.}
    \end{subfigure}%
    \hfill
    \begin{subfigure}{0.45\textwidth}
        \centering
        \includegraphics[width=\textwidth]{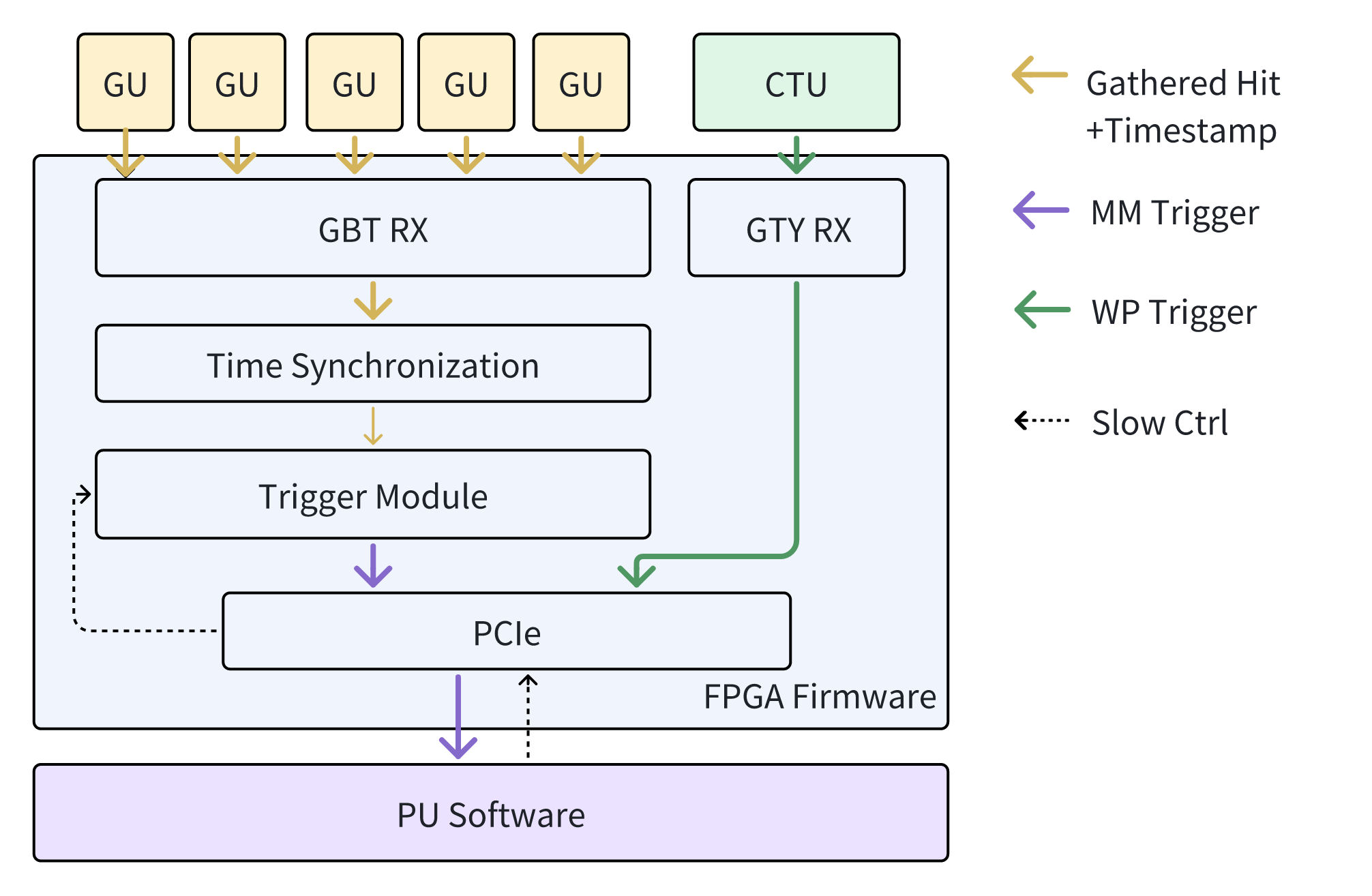}
        \caption{Illustration of the PU firmware.}
    \end{subfigure}
    \caption{Structure of the GU and PU firmware.}
    \label{fig:firmware_structure}
\end{figure}

The MM trigger's GUs adopt a hardware architecture closely resembling that of the CTU~\cite{JUNO_Initial_Performance_2025}, but their internal firmware is completely different. In total, five GUs collect trigger hit signals from 134 BECs and forward the processed data to the PU as shown in Figure~\ref{fig:firmware_structure}(a). Each GU integrates a White Rabbit (WR) timing system~\cite{lipinski2011white}, which provides timestamps with an 8~ns resolution to ensure precise synchronization among all PMTs. 
Communication between the GU and the BECs relies on the GigaBit Transceiver protocol~\cite{Marin2015GBT}. This protocol is specifically chosen because it ensures a fixed and deterministic data transmission latency, which is essential for real-time trigger synchronization. It supports a bandwidth of 8~Gbps by transmitting 128 bits per 62.5~MHz clock cycle. In addition, the GU firmware incorporates the IPbus protocol~\cite{ghabrous2015ipbus}, allowing remote configuration and firmware updates via the WR network. 
To facilitate the likelihood algorithm (detailed in Section~\ref{sec:Trigger_Algorithms}), the GU firmware pre-aggregates the incoming PMT hits into 16 spatial groups, which directly correspond to the $4 \times 4$ spatial bins used in the likelihood calculation. The aggregated hit information, paired with precise WR timestamps, is then transmitted to the PU via the GBT protocol for the final trigger decision.

Upon receiving the trigger hit information from the five GUs, the PU first synchronizes the incoming data streams based on their timestamps. This alignment is necessary because the GUs are distributed across two different electronics rooms with varying optical fiber lengths. Once synchronized, the raw 16~ns hit data are integrated into 48~ns time bins. Four consecutive 48~ns bins are then grouped to form a 192~ns time window, which is continuously fed into the fast likelihood-based trigger algorithm (Figure~\ref{fig:firmware_structure}(b)). This algorithm evaluates the spatial and temporal clustering of hits. A trigger, containing the trigger timestamp and the number of hits, is issued to the host computer when the likelihood exceeds a configurable threshold. The PU firmware receives a WP trigger signal from the CTU via a high-speed GTY transceiver link, operating at 5~Gbps with a custom 8b/10b-encoded protocol. This signal is then forwarded to the PU software shown in Figure~\ref{fig:PU}(b), serving as a muon veto for the Transient Neutrino Burst Monitor.

\section{Reconstruction Algorithm}
\label{appendix:reco}

The vertex reconstruction used in this analysis is a time-based algorithm that determines the event vertex $\mathbf{r}=(x,y,z)$ and the emission time $t_{0}$ from the first t-q (time and charge) pair signals recorded by the PMTs. The method is based on VTREP\cite{JUNO_ReactorOscillation_2025} framework, which is a timing-based vertex reconstruction algorithm, with optimizations tailored for MM low-energy events. For a trial vertex, the expected photon time-of-flight (TOF) to PMT $i$ is
\begin{equation}
    t^{\mathrm{TOF}}_{i}
    = \frac{n_{\mathrm{eff}}}{c}\,
      \left|\mathbf{r}-\mathbf{r}_{i}\right|,
\end{equation}
where $n_{\mathrm{eff}}$ is the effective refractive index and  $\mathbf{r}_{i}$ is the PMT position. The time residual is defined as
\begin{equation}
    t^{\mathrm{res}}_{i}
    = t^{\mathrm{FHT}}_{i} - \left(t_{0} + t^{\mathrm{TOF}}_{i}\right),
\end{equation}
where $t^{\mathrm{FHT}}_{i}$ represents the measured first hit time (FHT) of the $i$-th PMT. The distribution of $\{t^{\mathrm{res}}_{i}\}$ is used to identify the dominant photon-arrival cluster.

The main reconstruction algorithm proceeds in two stages. First, a coarse radial scan is performed along the direction of the initial seed vertex to identify a region whose residual-time pattern is most compatible with the observed early photon population.   Second, a multivariate minimization refines $(t_{0},\mathbf{r})$ using a likelihood function that quantifies the agreement between measured and predicted PMT times. The likelihood is constructed from the PMT timing response and modeled as
\begin{equation}
    \mathcal{L}(t_{0},\mathbf{r})
    = \prod_{i}
      \exp\!\left[
         -\frac{\left(t^{\mathrm{res}}_{i}\right)^{2}}
                {2\,\sigma_{i}^{2}}
      \right],
\end{equation}
where $\sigma_{i}$ is the timing resolution parameter. To handle the timing asymmetry, a bifurcated $\sigma_{i}$ is used: $\sigma_{\mathrm{early}}$ (dominated by PMT transit-time spread) for $t^{\mathrm{res}}_{i} < 0$, and a larger $\sigma_{\mathrm{late}}$ for $t^{\mathrm{res}}_{i} \ge 0$ to account for scattering and scintillation decay tails.
The final vertex estimate is obtained by maximizing $\mathcal{L}$, or equivalently minimizing $-\ln\mathcal{L}$:
\begin{equation}
    (\hat{t}_{0},\hat{\mathbf{r}})
    = \arg\min_{t_{0},\mathbf{r}}
      \left[-\ln \mathcal{L}(t_{0},\mathbf{r})\right].
\end{equation}

This vertex reconstruction algorithm operates well for events in a wide-range-energy region, including MM low energy events, with enough accuracy and fast reconstruction speed. According to MC estimation, the spatial resolutions in x, y and z directions are 1.1~m each for 40--60~keV gamma, 0.6~m each for 110--130~keV gamma, which meet the analysis requirements for $^{14}$C.

% \section{Bayesian Blocks Algorithm}
% \label{appendix:BBA}

\section{Bayesian Blocks Algorithm}
\label{appendix:BBA}

The Bayesian Blocks Algorithm (BBA) is a non-parametric and model-independent
method for detecting statistically significant variations in the rate of events~\cite{Scargle1998, Scargle2013}. 
Unlike conventional methods that require fixed-size time bins or assume a
specific signal time profile, the BBA adaptively partitions the time axis into
segments---referred to as \emph{blocks}---such that the event rate within each
block is statistically consistent with a constant value. This feature makes the
algorithm particularly well suited for identifying transient or burst-like
signals of arbitrary duration and shape, such as those expected from a
supernova neutrino burst.

\subsection{Mathematical Framework}

Given a sequence of event times $\{ t_i \}$, the algorithm seeks the optimal
segmentation that maximizes a fitness function constructed from the Poisson
likelihood of each block. For a block spanning a time interval $\Delta t$ that
contains $N$ events, the corresponding fitness function is defined using the
Cash statistic~\cite{Cash1979}:

\[
\mathcal{F}(N, \Delta t) = N \left[ \ln \left( \frac{N}{\Delta t} \right) - 1 \right] ,
\]

where the first term corresponds to the Poisson log-likelihood of a constant
event rate, and the second term ensures proper normalization. The total fitness
of the segmentation is the sum of the fitness values of all blocks.

To prevent overfitting---i.e., the creation of excessively many
blocks---a prior penalty is introduced. The optimal division points, called \emph{change points}, are obtained via dynamic programming by maximizing the sum of block fitness minus the prior term.

\subsection{Application in the MM Trigger System}

In the Transient Neutrino Burst Monitor described in 
Section~\ref{sec:Supernova_Monitor}, each trigger provides a timestamp and an
associated number of trigger hits (Trigger $N_{\mathrm{hit}}$). After applying the
high-energy selection, muon veto, and IBD delayed-coincidence requirement, the
remaining triggers define a time series dominated by residual background
events. Detecting a sudden rise in the event rate can indicate the onset of a
supernova burst.

The BBA is applied to this time series in real time. Under the
\emph{null hypothesis} ($H_0$), the interval is modeled as a single block with
a constant background rate. Under the \emph{burst hypothesis} ($H_1$), the data
favor a segmentation with at least one additional block whose rate is higher
than the background. Comparing the likelihoods of these hypotheses yields a
test statistic:

\[
\mathrm{TS}_{\mathrm{high}} =
\begin{cases}
-2 \ln (L_0 / L_1), & L_1 > L_0, \\
0, & L_1 \le L_0,
\end{cases}
\]

where $L_0$ and $L_1$ denote the Poisson likelihoods under the single-block and
multi-block hypotheses, respectively. A non-zero
$\mathrm{TS}_{\mathrm{high}}$ indicates the detection of a statistically
significant increase in the event rate.

\bibliographystyle{unsrt}
\bibliography{reference}

\end{document}

%% file: Authors_v2.tex
\author[38]{Thomas Adam}
\author[14]{Fengpeng An}
\author[65]{Costas Andreopoulos}
\author[48]{Giuseppe Andronico}
\author[60]{Nikolay Anfimov}
\author[50]{Vito Antonelli}
\author[60]{Tatiana Antoshkina}
\author[38]{Jo\~{a}o Pedro Athayde Marcondes de Andr\'{e}}
\author[36]{Didier Auguste}
\author[60]{Nikita Balashov}

% ...
\affil[14]{Sun Yat-Sen University, Guangzhou, China}
% ...
\affil[36]{IJCLab, Universit\'{e} Paris-Saclay, CNRS/IN2P3, 91405 Orsay, France}
% ...
\affil[38]{IPHC, Universit\'{e} de Strasbourg, CNRS/IN2P3, F-67037 Strasbourg, France}
% ...
\affil[48]{INFN Catania and Dipartimento di Fisica e Astronomia dell Universit\`{a} di Catania, Catania, Italy}
% ...
\affil[50]{INFN Sezione di Milano and Dipartimento di Fisica dell Universit\`{a} di Milano, Milano, Italy}
% ...
\affil[60]{Joint Institute for Nuclear Research, Dubna, Russia}
% ...
\affil[65]{The University of Liverpool, Department of Physics, Oliver Lodge Laboratory, Oxford Str., Liverpool L69 7ZE, UK, United Kingdom}
% ...
